\documentclass[showpacs,showkeys,11pt,
preprint,preprintnumbers,nofootinbib,
groupedaddress,superscriptaddress,amsmath,amssymb]{revtex4}
\usepackage{amsfonts}
\usepackage{amsmath}
\usepackage{graphicx,subfigure}
\usepackage{caption}
\usepackage{float}
\usepackage{rotating}
\usepackage[export]{adjustbox}
\usepackage{epsfig}
\usepackage{url}
\usepackage{multirow}
\usepackage{feynmp}

\newcommand {\be}{\begin{equation}}
\newcommand {\ee}{\end{equation}}
\newcommand {\ba}{\begin{eqnarray}}
\newcommand {\ea}{\end{eqnarray}}

\begin{document}
\title{Probing triple higgs self coupling and effect of beam polarization in lepton colliders}

\author{Ijaz Ahmed}
\author{Ujala Nawaz}
\email{Ijaz.ahmed@cern.ch}
\affiliation{Riphah International University, Sector I-14, Hajj Complex, Islamabad Pakistan}
\author{Taimoor Khurshid}
\email{taimoor.khurshid@iiu.edu.pk}
\affiliation{International Islamic University, H-10, Islamabad}
\author{Shamona Fawad Qazi}
\email{shamona@qau.edu.pk}
\affiliation{Quaid-i-Azam University, Islamabad}


\begin{abstract}
One of the main objectives of almost all future (lepton) colliders is to measure the self-coupling of triple Higgs in the Standard Model. By elongating the Standard Model's scalar sector, using incipient Higgs doublet along with a quadratic (Higgs) potential should reveal many incipient features of the model and the possibility of the emergence of additional Higgs self-couplings. The self-coupling of the Higgs boson helps in reconstructing the scalar potential. The main objective of this paper is to extract Higgs self-coupling by numerically analyzing several scattering processes governed by two Higgs doublet models (2HDM). These scattering processes include various possible combinations of final states in the triple Higgs sector. The determination of production cross-section of scattering processes is carried out using two different scenarios, with and without polarization of incoming beam and is extended to a center of mass energy up to $\sqrt{s}$ = 3 $TeV$. The computation is carried out in Type-1 2HDM. Here we consider the case of exact alignment limit ($s_{\beta \alpha}$ = 1) and masses of extra Higgs state are equal that is $m_H = m_{H^0} = m_{A^0} = m_{H^\pm}$. This choice is made to minimize the oblique parameters. The decays of the final state of each process are investigated to estimate the number of events at integrated luminosity of $1ab^{-1}$ and $3ab^{-1}$.
\end{abstract}
\maketitle
\section{Introduction}
It is the Higgs mechanism that explains the origin of mass for elementary particles in Standard Model (SM) via an electroweakly broken symmetry mechanism. After the discovery of the most awaited Higgs particle, having a mass of 125 $GeV$, by CMS and ATLAS experiments, one of the major aims of the Large Hadron Collider (LHC) \cite{lab1,lab2} was to study the properties of this particle such as precision measurement of its mass, production rate, coupling to SM particles, spin-parity, etc. These studies indicate that the discovered particle is in perfect agreement with the SM predictions. The Higgs sector could be more complicated for what is understood as yet and Beyond the Standard Model (BSM) effects could appear from exact measurements of the couplings with fermions and bosons, which can be determined from the Higgs boson production processes and decays. Two Higgs doublet model (2HDM) is one of the simplest extensions of the Standard Model. A second complex doublet SU(2) is added into 2HDM, as a result, we end up having five Higgs states in 2HDM, one CP-odd scalar ($A^0$), two CP even scalars ($H^0$, $h^0$), and two charged Higgs bosons ($H^+$, $H^-$).\\
LHC is the biggest particle accelerator, where two beams of protons collide with each other and the resulting collision events are recorded. These events give us information about the beginning of the universe and properties of particles which make up the universe. Hadron colliders are discovery machines as they can reach the highest possible beam energies and therefore, are powerful probes to new energy ranges. For examining the Higgs particles and their properties a lepton collider machine is necessary, as these are the natural precision machines. In lepton collider, the initial states of each event are known accurately and high precision of measurements is possible to achieve. The future circular collider (FCC-ee) \cite{lab3} at CERN, the circular electron-positron collider (CEPC) \cite{lab4} in china and the international linear collider (ILC) \cite{ilc} is designed to produce many Higgs bosons and investigate their properties.\\
The measurement of Higgs self-coupling gives the information about the EWSB mechanism and understanding of the scalar potential of the Higgs field. The process $e^+ e^- \rightarrow ZHH$ which is also known as Higgs-strahlung is best suited to calculate the trilinear Higgs self-coupling in the Standard Model. A similar effort was made earlier where trilinear Higgs couplings with various Higgs bosons pairs connected with the Z boson were examined \cite{lab5}. Some of the double and triple Higgs boson couplings were also examined in \cite{lab6,lab7}. Also trilinear and quartic Higgs coupling have been studied in the past within MSSM \cite{lab8}.\\
In our study, several scattering processes and their possible Higgs self-couplings are determined. The production cross-section as a function of the center of mass-energy and polarization of incoming beam is calculated. The results are obtained within the framework of 2HDM taking into consideration the theoretical and experimental constraints. 


\section{Two Higgs Doublet Model 2HDM}
The simplest extension to SM is 2HDM with a different Higgs field but based on an identical gauge field with the same fermion content. The 2HDM consists of 2 Higgs isospin doublets containing hypercharge content of original Higgs field, with 8 degrees of freedom. 
\begin{align*}
\phi_1 = 
\begin{pmatrix}
\phi_1^+ \\
\phi_1^0 \\
\end{pmatrix}, 
\phi_2 =
\begin{pmatrix}	
 \phi_2^+ \\
 \phi_2^0
 \end{pmatrix}
\end{align*}    
When symmetry is spontaneously broken, in addition to three gauge bosons, the $W^{\pm}$ and $Z^0$, we get five new physical Higgs bosons: the two CP-even neutrals h and H, one CP-odd neutral, and two charged scalars $H^{\pm}$.\\
When a discrete $Z_2$ symmetry is applied on the Lagrangian, it results in four possible types of 2HDM which satisfy the GWP \cite{lab37} criterion. \\
In type 1, both quarks and leptons couple to $\phi_2$ while in type 2, up type quarks couple to $\phi_2$, whereas down type quarks and charged leptons couple to $\phi_1$. Similarly, in type 3, up type quarks and charged leptons couple to $\phi_2$ while down type quarks couple to $\phi_1$. This type is sometimes called flipped. In type 4, all charged leptons couple to $\phi_1$ while all quarks couple to $\phi_2$. This type is also called lepton specific.

\subsection{Softly Broken $Z_2$  Symmetry}
A discrete $Z_2$ symmetry is applied on the Lagrangian which constrain it. The Higgs basis is defined as
\begin{align*}
 \phi_1 = 
 \begin{pmatrix}
 G^+ \\
 \frac{1}{\sqrt{2}} [\nu + S_1 + \iota G^0] \\
 \end{pmatrix},
 \phi_2 = 
 \begin{pmatrix}
 H^+ \\
 \frac{1}{\sqrt{2}} [S_2 + \iota S_3] 
 \end{pmatrix}
 \label{eq:4.1}
\end{align*}

where $\quad (\iota=1,2)$ and $\nu= \sqrt{\nu^{2}_1 + \nu^{2}_2}$. The hermitian Klein-Gordon fields are $G^0$, $S_1$, $S_2$ and $S_3$  while complex Klein-Gordon fields are $G^+$ and $H^+$.\\
Invoking a $Z_2$ symmetry removes flavor changing neutrals current (FCNCs). Therefore in the Higgs basis, the scalar potential is written as

\begin{align}
V(\phi_1, \phi_2)= m^2_1 |\phi_1|^2 + m^2_2 |\phi_2|^2 - [m^2_3 \phi_1^+ \phi_2 + h.c.] + 
 \frac{\Lambda_1}{2} \Bigg| (\phi_1^+ \phi_1)^2 + \frac{\Lambda_2}{2} (\phi_2^+ \phi_2)^2 + \Lambda_3 |\phi_1|^2 |\phi_2|^2 + \Lambda_4 |\phi_1^+ \phi_2| \Bigg|^2 \\ + \Bigg[ \frac{\Lambda_5}{2}(\phi_1^+ \phi_1)^2 + h.c. \Bigg] + [(\Lambda_6 \phi_1^+ \phi_1 + \Lambda_7 \phi_2^+ \phi_2) \phi_1^+ \phi_2 + h.c.],
\label{eq:4.2}		
\end{align}

The coupling constant $m^2_1$, $m^2_2$, $\Lambda_1$, $\Lambda_2$, $\Lambda_3$ and $\Lambda_4$ are real and the complex parameters are $m^2_3$, $\Lambda_5$, $\Lambda_6$ and $\Lambda_7$ but for simplification they are taken as real. The scalar potential can be decomposed as a sum of quadratic, cubic and quartic interactions. The quadratic terms define the physical Higgs state and its masses. Diagonalization of the quadratic mass terms gives masses of all extra Higgs bosons \cite{lab38}. The $\sin{(\beta - \alpha)}$ is the mixing angle among the CP-even Higgs state and is also denoted by $s_{\beta \alpha}$. \\   

The exact alignment limit i.e. $s_{\beta \alpha}$ = 1 is considered, so that $h^0$ become indistinguishable from the Standard Model Higgs boson with respect to coupling and mass. Hence there are only six independent parameters of the model which are important for our study, which include $m_h$, $m_{H^0}$, $m_{A^0}$, the ratio of vacuum expectation value $\tan{\beta}$, $s_{\beta \alpha}$ and ${m_3}^2$. The parameter ${m_3}^2$  indicates that, how is the discrete symmetry broken \cite{lab39}. In Equation ~\ref{eq:4.2} the cubic and quartic terms define the interactions and couplings between the new states in 2HDM.

\subsection{Theoretical And Experimental Constraints}

The parameters of scalar potential in 2HDM are reduced both by the theoretical developments, as well as results of experimental searches. The theoretical constraints to which 2HDM is subjected, comprise of vacuum stability, unitarity, and perturbativity.
\begin{itemize}
\item \textbf{Stability}: Requirement of a stable vacuum keeps the potential bounded from below. The potential must be positive at large values of fields for any direction in the plane \cite{lab40,lab41}.
\item \textbf{Unitarity}: Unitarity constraints do not allow the scattering amplitudes to have probability more than unity. The amplitude needs to be flat at asymptotically large values of energy \cite{lab42,lab43}.
\item \textbf{Perturbitivity}: The potential needs to be perturbative to fulfill the requirement that all quartic couplings of scalar potential obey $\Lambda_i \leq 8 \pi$ for all $i$ 
\end{itemize}
With the help of 2HDMC-1.7.0 \cite{lab44}, the parameters are tested as if, they obey the above mentioned constraints. The 2HDM parameters are also constrained by experimental researches. According to a recent study \cite{lab45}, it is observed that  flavor limits are present and Figure 3 in there shows the available region which is not excluded even today. The charged Higgs $H^{\pm}$ present in 2HDM, which is comparable to SM, also makes a novel contribution in flavor limits. According to current experimental results at LHC (give reference) and (give reference), masses of all extra Higgs bosons are set to be equal that is $m_H=m_{H^0}m_{A^0}=m_{H^{\pm}}$. This choice minimizes the oblique parameters and all the electroweak observables are close to SM. The decay of $H^0$ to vector boson pair is suppressed in the exact alignment limit. According to \cite{lab43} in type-1, the neutral meson mixing and results of $B^0_s \rightarrow \mu^+ \mu^-$ restrict the low $\tan{\beta}$. The analysis is therefore, carried out in range 2 $< \tan{\beta} <$ 40. The regions in which theoretical constraints are obeyed by ${m_3}^2$ are measured with the help of 2HDMC.

\begin{figure}[h]
 \centering
\includegraphics[scale=0.60]{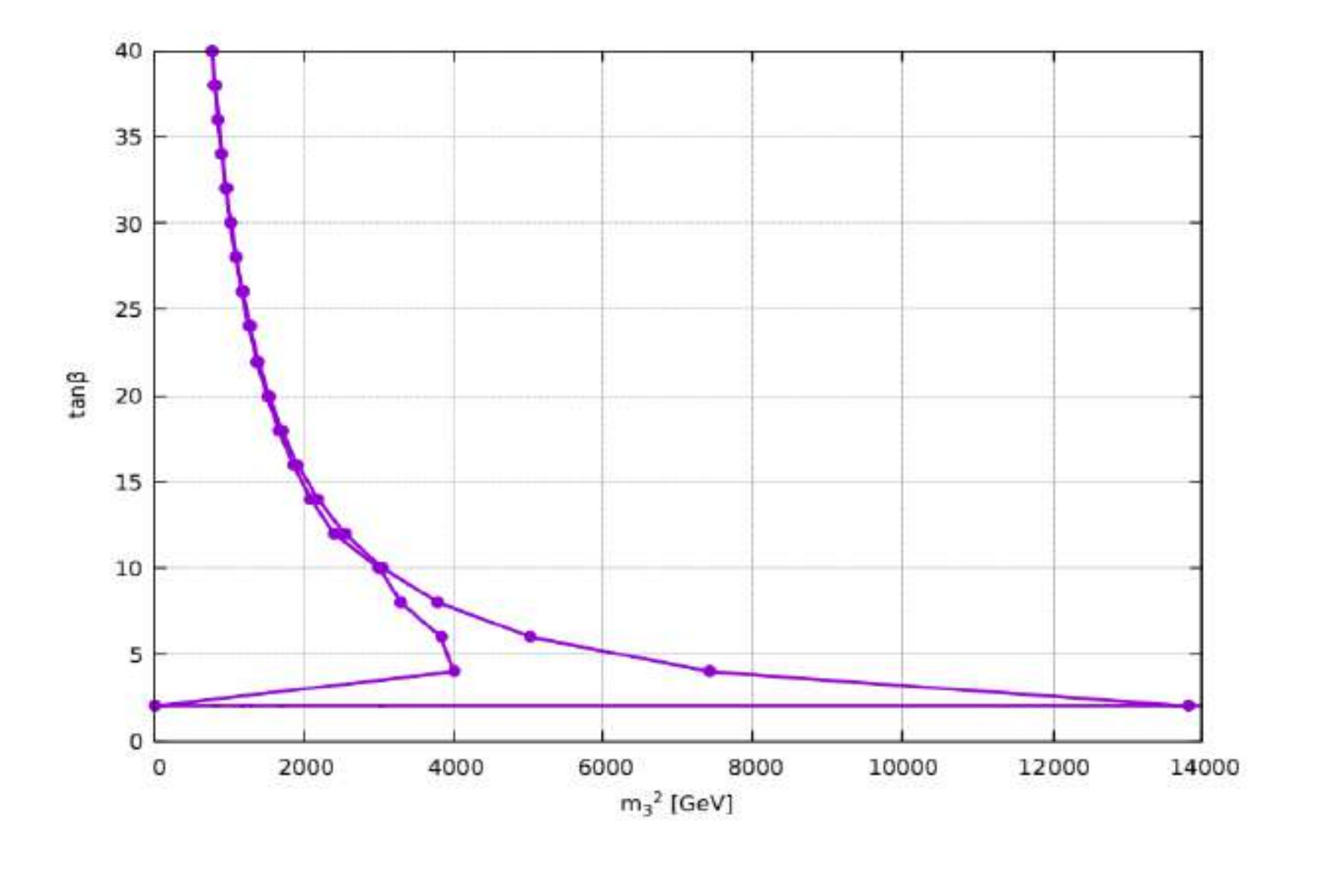}
    \caption{The plot between soft symmetry breaking term versus $\tan{\beta}$. The region under bounded curves shows the theoretical constraints obeyed.}
    \label{fig:1}
\end{figure}

\begin{table}[h!]
\begin{center}
 \begin{tabular}{|c|c|c|c|c|c|}
\hline
Benchmark & Yukawa Type & $m_h$ $[GeV]$ & $m_{H} = m_{A} = m_H^{+}$ [GeV]  & $s_{\beta \alpha}$ & $t_{\beta}$ \\
\hline
1 & Type-I & 125 & 175 & 1 & 2.40 \\
\hline
\end{tabular}%
\caption{ The  range of independent parameters calculated within Type-1 2HDM}
\label{table:1}
\end{center}
\end{table}
\subsection{Higgs Self-couplings in 2HDM}
In 2HDM, Higgs self-couplings as a function of $\Lambda_i$ are given by Equations ~\ref{eq:4.3} to ~\ref{eq:4.8}. The values for two independent parameters are selected to be $s_{\beta \alpha}$ = 1 and $c_{\beta \alpha}$ = 0. Due to exact alignment limit and equal masses of all  extra Higgs bosons, three other parameters $\Lambda_4$, $\Lambda_5$ and $\Lambda_6$ also vanish. Among all the Higgs self couplings, only the one given by $g_{hhH}$  vanishes. However the couplings $g_{hHH}$ and $g_{hAA}$ are equal to each other and $g_{HHH}/g_{HAA}$ = 3. These predictions can also be checked experimentally. If $\Lambda_{345}= \Lambda_3 + \Lambda_4+ \Lambda_5$, then we can write
\setlength{\belowdisplayskip}{0pt} \setlength{\belowdisplayshortskip}{0pt}
\setlength{\abovedisplayskip}{0pt} \setlength{\abovedisplayshortskip}{0pt}

\begin{equation}
g_{h^0 h^0 h^0}= -3 \iota \nu (( \Lambda_7 c^2_{\beta \alpha} + 3\Lambda_6 s^2_{\beta \alpha}) c_{\beta \alpha} + (\Lambda_{345} c^2_{\beta \alpha} + \Lambda_1 \delta^2_{\beta\alpha})s_{\beta \alpha} ) \underset {c\beta\alpha \rightarrow 0}{=} -3 \iota \nu \Lambda_1
\label{eq:4.3}
\end{equation}
 
\begin{equation}
g_{h^0 h^0 H^0} = - \iota \nu (( \Lambda_{345} (1-3s^2_{\beta\alpha}) + 3\Lambda_1 s^2_{\beta\alpha})c_{\beta\alpha} + 3(\Lambda_6(2-3s^2_{\beta\alpha})- \Lambda_7 c^2_{\beta\alpha})s_{\beta\alpha}) \underset {c\beta\alpha \rightarrow 0}{=} 0
\label{eq:4.4}
\end{equation}

\begin{equation}
g_{h^0 H^0 H^0}= - \iota \nu ((3 \Lambda_1 c^2 _{\beta\alpha} + \Lambda_{345} (3s^2_{\beta\alpha}-2))s_{\beta\alpha} + 3(\Lambda_6 + \Lambda_7 \% s^2_{\beta\alpha} - 3 \Lambda_6 s^2_{\beta\alpha})c_{\beta\alpha}) \underset {c\beta\alpha \rightarrow 0}{=} - \iota \nu \Lambda_3
 \label{eq:4.5}
\end{equation}

\begin{equation}
g_{h^0 A^0 A^0}= -\iota \nu (\Lambda_7 c_{\beta\alpha} + (\Lambda_3 + \Lambda_4 - \Lambda_5)s_{\beta\alpha}) \underset {c\beta\alpha \rightarrow 0}{=} - \iota \nu \Lambda_3
\label{eq:4.6} 
\end{equation}

\begin{equation}
g_{H^0 H^0 H^0}= -3 \iota \nu ((\Lambda_1 c^2_{\beta\alpha} + \Lambda_{345}s^2_{\beta\alpha})c_{\beta\alpha} - \Lambda_7 s^2_{\beta\alpha} -3\Lambda_6 c^2_{\beta\alpha})s_{\beta\alpha} \underset {c\beta\alpha \rightarrow 0}{=} 3\iota \nu \Lambda_7
\label{eq:4.7}
\end{equation}

\begin{equation}
g_{H^0 A^0 A^0}= - \iota \nu ((\Lambda_3 + \Lambda_4 - \Lambda_5)c_{\beta\alpha} - \Lambda_7 s_{\beta\alpha}) \underset {c\beta\alpha \rightarrow 0}{=}  \iota \nu \Lambda_7
\label{eq:4.8}
\end{equation}

\section{Tripple Higgs Self Coupling and Production Cross-section}

It is normally believed that ILC will perform efficiently in precision measurements as compared to LHC, due to it's clean environment, fixed center-of-mass energy, and attainability of polarised beam. This fact emphasizes the importance of ILC for Higgs sector in terms of calculations of different scattering processes, their self-couplings, branching ratios and estimation of number of events.


\subsubsection{Measurement of Higgs Self-Coupling}

\begin{figure}[h]
\centering
\includegraphics[scale=0.50]{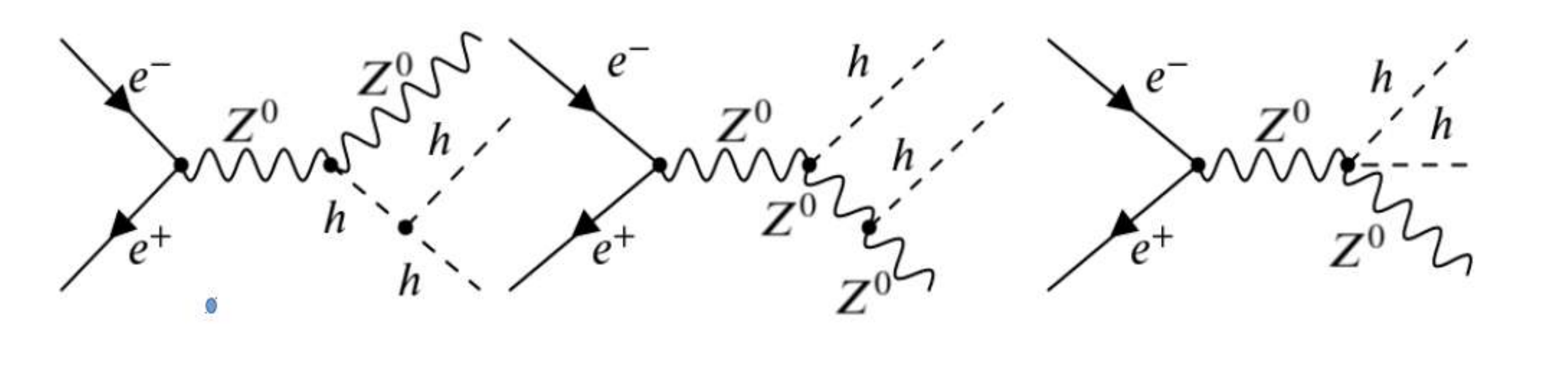}
\caption{The left-most diagram contains the Higgs self coupling vertex corresponding to the process $e^+ e^- \rightarrow Zhh$.}
 \label{fig:2_5.1}
\end{figure}

The trilinear self-coupling can be measured directly or indirectly by using the Higgs-boson-pair production cross section, or through the measurement of single-Higgs-boson production and decay modes, respectively. In fact, the Higgs-decay partial widths and the cross sections of the main single-Higgs production processes depend on the Higgs-boson self-coupling via weak loops, at next-to-leading order in electroweak interaction. Let's consider the scattering process $e^- e^+ \rightarrow Zhh$ to study the trilinear coupling of the Higgs boson. In SM the Feynman diagrams of this process are shown in Figure ~\ref{fig:2_5.1}.

The measurement of Higgs self-coupling within 2HDM is difficult due to the presences of more than one Higgs. The couplings in which $h^0$, $H^0$, and $A^0$ are intermediated, do not make a noticeable contribution because of their absolute value, which is less than $10^{-6}$, so they can be neglected. Significant contributions are found to be from $Z^0$ coupling only, that's why only those Feynman diagrams are taken into account in which Z boson is intermediated. The scattering processes with various combinations of trilinear Higgs self-couplings need to be considered, i.e. $ZHh$, $ZAh$,  $HHH$, $hhh$, $Ahh$, $AAh$, $AAH$, $Hhh$ and $Hhh$.

\begin{figure}[h]
    \centering
\includegraphics[scale=0.50]{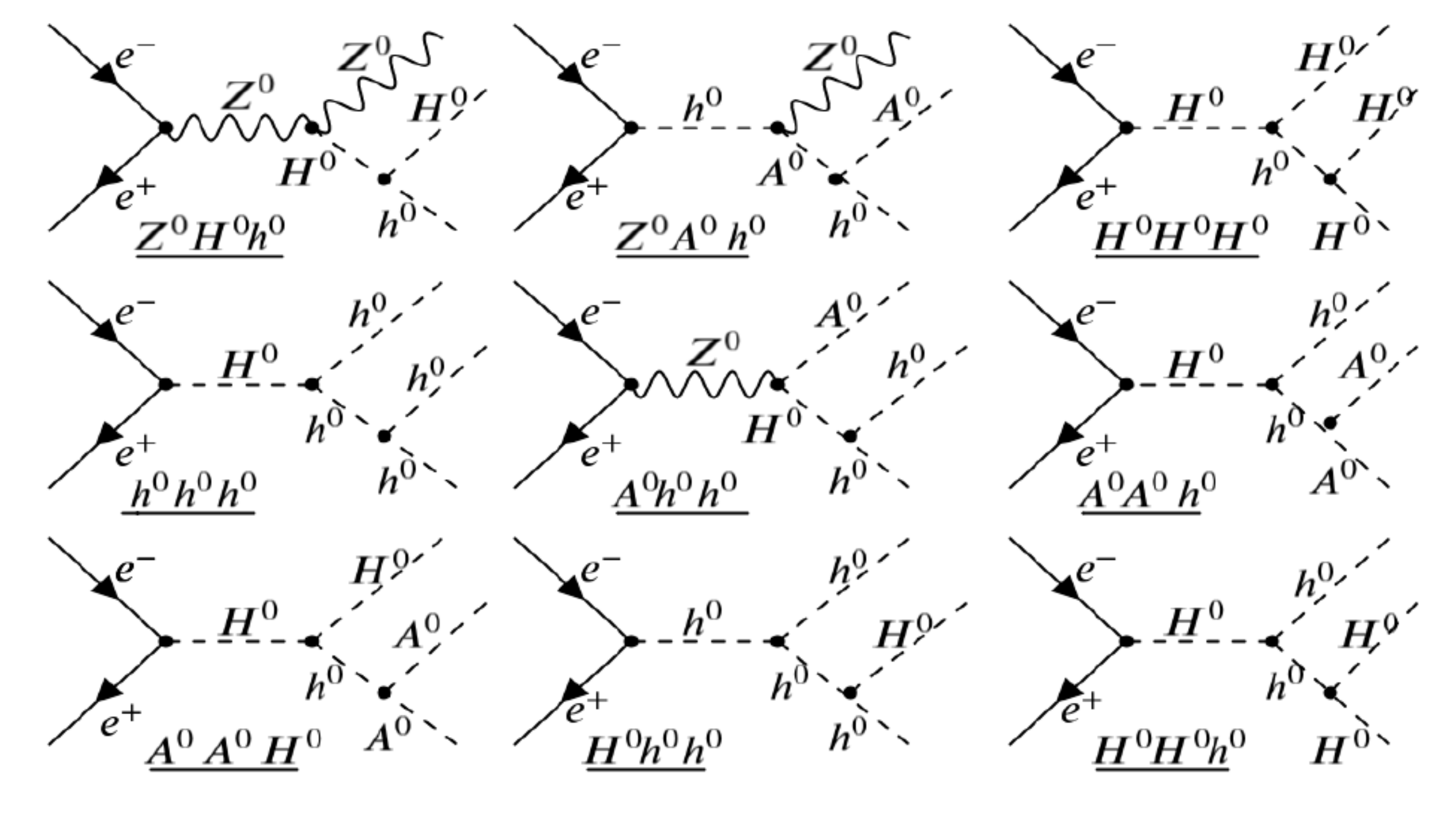}
    \caption{The feynman diagram of various processes in 2HDM. Their amplitude shows that their cross-section is less than $10^{-11}$ pb, so their self coupling could be neglected.}
    \label{fig:3_5.2}
\end{figure}
The Feynman diagrams of the possible processes are shown in Figure ~\ref{fig:3_5.2}. Their cross-section is less than $10^{-11}$pb, therefore it is not possible to detect them and can be easily neglected.

\begin{table}[h]
\begin{center}
 \begin{tabular}{|c|c|}
\hline
Scatering Processes & Higgs self-coupling \\
\hline
$e^- e^+ \rightarrow Z^0 A^0 A^0$ & $g_{h^0 A^0 A^0}$ \\
$e^- e^+ \rightarrow Z^0 H^0 H^0$ & $g_{h^0 H^0 H^0}$ \\
$e^- e^+ \rightarrow Z^0 h^0 h^0$ & $g_{h^0 h^0 h^0}$ \\
$e^- e^+ \rightarrow A^0 H^0 h^0$ & $g_{H^0 A^0 A^0}$ \\
$e^- e^+ \rightarrow A^0 H^0 H^0$ & $g_{H^0 A^0 A^0}$, $g_{H^0 H^0 H^0}$ \\
\hline
\end{tabular}%
\caption{ The tripple Higgs self-coupling contributing to scattering process while using the exact alignment limit equal to one and extra Higgs masses to be equal.}
\label{table:2}
\end{center}
\end{table}
To calculate the Higgs self-coupling in two Higgs doublet model, we consider the second set of scattering processes shown in Table ~\ref{table:2}. These scattering processes are the only ones which can give the cross-section greater than attobarn. In Equation ~\ref{eq:4.4b} $g_{h^0 h^0 H^0}$ approaches zero so this coupling vanishes. The cross-section of scattering process $e^- e^+ \rightarrow ZAA$ makes it possible to determine the coupling $g_{h^0 A^0 A^0}$. The coupling $g_{h^0 H^0 H^0}$ can be determined by measuring the cross-section of process  $e^- e^+ \rightarrow ZHH$. The cross-section of $e^- e^+ \rightarrow Zhh$ extracts the coupling  $g_{h^0 h^0 h^0}$ which could be the same as determined in SM. The coupling $g_{H^0 A^0 A^0}$ can be determined by two processes, $e^- e^+ \rightarrow AHh$ and $e^- e^+ \rightarrow AHH$, whereas the last mentioned process can also give $g_{H^0 H^0 H^0}$. 

\section{The production cross-section of scattering processes}

For the computation of production cross-section of various scattering processes, CalcHEP$\_$3.7.6 package \cite{lab46} is used. The parameters of Standard Model are used from the \cite{lab47}, which are $m_e$ = 0.51099 $MeV$, $m_Z$ = 91.1876 $GeV$ and $s_w$ = 0.474. The Higgs boson mass is taken to be $m_h$ = 125.09 $GeV$. The 2HDM parameters are already discussed in previous chapter. The cross-sections of various scattering process are given below:
\subsubsection{$e^- e^+ \rightarrow Zhh $}
\begin{figure}
\centering     
\subfigure[]{\label{fig:a}\includegraphics[width=80mm]{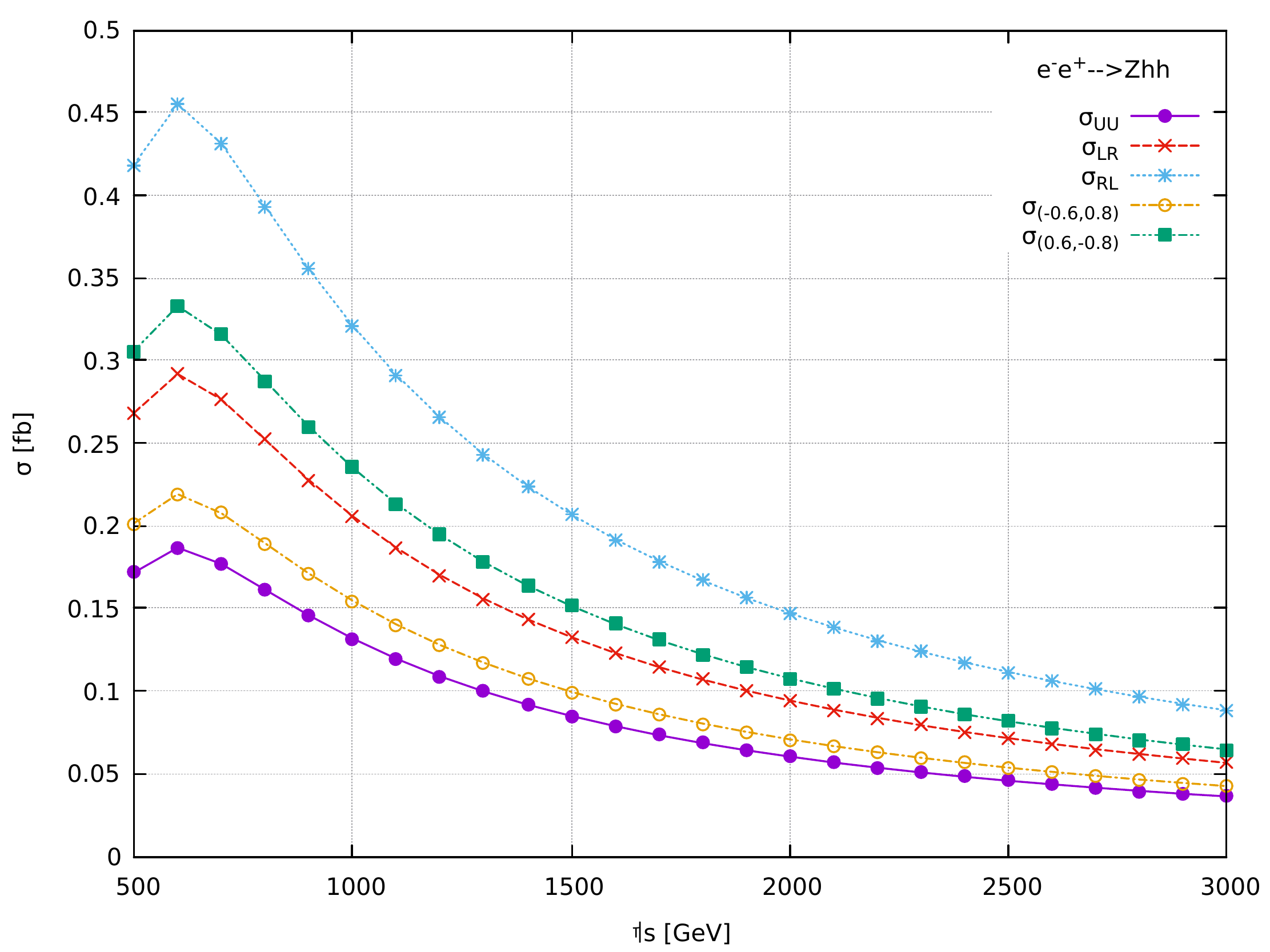}} 
\caption{Total cross-section $\sigma$ (fb) as a function of $\sqrt{s}$ (GeV) at Higgs masses $m_H$ = 150 $GeV$ or $m_H$ = 175 $GeV$, $\tan{\beta}$ = 10 and $s_{\beta \alpha}$=1 are shown.} 
\label{fig:4}
\end{figure}
The cross-section of this prominent channel is shown in the Figure ~\ref{fig:4}. It can be seen that the maximum cross-section of the unpolarized beam is approximately 0.187 $fb$ at 0.6 $TeV$, and decreases for higher energies. However the polarized beam enhances the cross-section as compared to the unpolarized beam. The right-handed electron beam and left-handed positron beam ($\sigma_{RL}$) maximize the cross-section which is around 0.455 $fb$ at 0.6 $TeV$. Moreover, the distribution of left-handed electron beam and right-handed positron beam ($\sigma_{LR}$), and two cases of polarization beam $\sigma_{(0.6,-0.8)}$ and $\sigma_{(-0.6,0.8)}$ are also given in Table ~\ref{table:3}. \\
\begin{table}[H]
\begin{center}
 \begin{tabular}{|c|c|c|c|}
\hline
  & 0.6 TeV   & 1 TeV & 3 TeV \\
\hline
  & $m_\phi$=150=175=300=500   & $m_\phi$=150=175=300=500 & $m_\phi$=150=175=300=500 \\
\hline
 $\sigma_{UU}$ & 0.187 $\pm$ 0.02 & 0.132 $\pm$ 0.0204 & 0.036 $\pm$ 0.01 \\
\hline
$\sigma_{LR}$ & 0.292 $\pm$ 0.031 & 0.206 $\pm$ 0.031 & 0.056 $\pm$ 0.02 \\
\hline 
$\sigma_{RL}$ & 0.455 $\pm$ 0.049 & 0.321 $\pm$ 0.050 & 0.082 $\pm$ 0.034 \\
\hline
$\sigma_{(0.6, -0.8)}$ & 0.332 $\pm$ 0.036 & 0.235 $\pm$ 0.036 & 0.065 $\pm$ 0.027 \\
\hline
$\sigma_{(-0.6, 0.8)}$ & 0.219 $\pm$ 0.023 & 0.155 $\pm$ 0.02 & 0.045 $\pm$ 0.018 \\
\hline
\end{tabular}%
\caption{ The cross-section in $fb$ with unpolarized and polarized incoming $e^- e^+$  beams. Where masses of extra Higgs states are varied over a range of  $m_H$ = 150 $GeV$ to $m_H$ = 500 $GeV$. In all cases, distribution of cross-section for both scenarios of with and without polarized beams is the same.}
\label{table:3}
\end{center}
\end{table}

\begin{table}[h!]
\centering
\resizebox{17cm}{!}{
\begin{tabular}{|c|c|c|c|c|c|c|c|c|c|c|c|c|c|c|c|c|}
\hline
  & \multicolumn{4}{| c |}{0.6 TeV}   & \multicolumn{4}{| c |}{1 TeV} & \multicolumn{4}{| c |}{3 TeV} \\
\hline
 &  $m_\phi$ =150 & $m_\phi$ =175 & $m_\phi$ =300 & $m_\phi$ =500 & $m_\phi$ =150 & $m_\phi$ =175 & $m_\phi$ =300 & $m_\phi$ =500 &$ m_\phi$ =150 & $m_\phi$ =175 & $m_\phi$ =300 & $m_\phi$ =500 \\
\hline
& \multicolumn{12}{| c |}{$e^{-} e^{+} \rightarrow AHh$}\\
\hline
$\sigma_{UU}$ & 0.0059 $\pm$ 0.0012 & 0.0023 $\pm$ 0.0004 & 0.0 $\pm$ 0.0 & 0.0 $\pm$ 0.0 & 0.0078 $\pm$ 0.0023 & 0.0059 $\pm$ 0.0016 & 0.0011 $\pm$ 0.0002 & 0.0 $\pm$ 0.0 & 0.0012 $\pm$ 0.0015 & 0.0010 $\pm$ 0.0014 & 0.0005 $\pm$ 0.0004 & 0.0002 $\pm$ 0.0001 \\
\hline 
$\sigma_{LR}$ & 0.0091 $\pm$ 0.0019 & 0.0037 $\pm$ 0.0007 & 0.0 $\pm$ 0.0 & 0.0 $\pm$ 0.0 & 0.0121 $\pm$ 0.0035 & 0.0093 $\pm$ 0.0025 & 0.00168 $\pm$ 0.0035 & 0.0 $\pm$ 0.0 & 0.0018 $\pm$ 0.0023 & 0.0016 $\pm$ 0.0018 & 0.00092 $\pm$ 0.00059 & 0.0004 $\pm$ 0.0002 \\
\hline
$\sigma_{RL}$ & 0.0143 $\pm$ 0.0029 & 0.0057 $\pm$ 0.0012 & 0.0 $\pm$0.0 & 0.0 $\pm$0.0 & 0.0189 $\pm$0.0056 & 0.0145 $\pm$ 0.0038 & 0.0026 $\pm$ 0.0006 & 0.0 $\pm$ 0.0 & 0.0029 $\pm$ 0.0038 & 0.0025 $\pm$ 0.0028 & 0.0014 $\pm$ 0.0009 & 0.00069 $\pm$ 0.0003 \\
\hline
$\sigma_{(0.6, -0.8)}$ & 0.0105 $\pm$ 0.0022 & 0.0042 $\pm$ 0.0008 & 0.0 $\pm$ 0.0 & 0.0 $\pm$ 0.0 & 0.0138 $\pm$ 0.0040 & 0.0186 $\pm$ 0.0049 & 0.0019 $\pm$ 0.0004 & 0.0 $\pm$ 0.0 & 0.0021 $\pm$ 0.0027 & 0.0018 $\pm$ 0.0021 & 0.0010 $\pm$ 0.0007 & 0.0005 $\pm$ 0.0002 \\
\hline
$\sigma_{(-0.6, 0.8)}$ & 0.0069 $\pm$ 0.0014 & 0.0027 $\pm$ 0.0006 & 0.0 $\pm$ 0.0 & 0.0 $\pm$ 0.0 & 0.0091 $\pm$ 0.0026 & 0.0070 $\pm$ 0.0018 & 0.0012 $\pm$ 0.0002 & 0.0 $\pm$ 0.0 & 0.0014 $\pm$ 0.0018 & 0.0012 $\pm$ 0.0014 & 0.0007 $\pm$ 0.0004 & 0.0003 $\pm$ 0.0001 \\
\hline
& \multicolumn{12}{| c |}{$e^- e^+ \rightarrow AHH$}\\
\hline
$\sigma_{UU}$ & 0.0018 $\pm$ 0.0004 & 0.0002 $\pm$ 0.00004 & 0.0 $\pm$ 0.0 & 0.0 $\pm$ 0.0 & 0.0042 $\pm$ 0.0013 & 0.0023 $\pm$ 0.0006 & 0.000014 $\pm$ 0.000003 & 0.0 $\pm$ 0.0 & 0.0009 $\pm$ 0.0011 & 0.0006 $\pm$ 0.0006 & 0.00017 $\pm$ 0.00087 & 0.000033 $\pm$ 0.0000094 \\
\hline
$\sigma_{LR}$ & 0.0029 $\pm$ 0.0006 & 0.00033 $\pm$ 0.00006 & 0.0 $\pm$ 0.0 & 0.0 $\pm$ 0.0 & 0.0065 $\pm$ 0.002 & 0.0035 $\pm$ 0.0009 & 0.00002152 $\pm$ 0.0000043 & 0.0 $\pm$ 0.0 & 0.0014 $\pm$ 0.0017 & 0.0010 $\pm$ 0.0009 & 0.00027 $\pm$ 0.00013 & 0.000051 $\pm$ 0.000017 \\
\hline
$\sigma_{RL}$ & 0.0046 $\pm$ 0.0010 & 0.00052 $\pm$ 0.0001 & 0.0 $\pm$ 0.0 & 0.0 $\pm$ 0.0 & 0.010 $\pm$ 0.003 & 0.0055 $\pm$ 0.0015 & 0.000033 $\pm$ 0.0000068 & 0.0 $\pm$ 0.0 & 0.0022 $\pm$ 0.002 & 0.0016 $\pm$ 0.0015 & 0.0004 $\pm$ 0.00020 & 0.000082 $\pm$ 0.000023 \\
\hline
$\sigma_{(0.6, -0.8)}$ & 0.0038 $\pm$ 0.0008 & 0.0003 $\pm$ 0.0008 & 0.0 $\pm$ 0.0 & 0.0 $\pm$ 0.0 & 0.0074 $\pm$ 0.0023 & 0.0040 $\pm$ 0.001 & 0.000024 $\pm$ 0.0000049 & 0.0 $\pm$ 0.0 & 0.0016 $\pm$ 0.0019 & 0.0011 $\pm$ 0.001 & 0.00031 $\pm$ 0.00015 & 0.0000602 $\pm$ 0.000017 \\
\hline
$\sigma_{(-0.6, 0.8)}$ & 0.0022 $\pm$ 0.0004 & 0.00025 $\pm$ 0.00005 & 0.0 $\pm$ 0.0 & 0.0 $\pm$ 0.0 & 0.0049 $\pm$ 0.0015 & 0.0026 $\pm$ 0.0007 & 0.000016 $\pm$ 0.0000033 & 0.0 $\pm$ 0.0 & 0.0010 $\pm$ 0.0012 & 0.00075 $\pm$ 0.0007 & 0.00020 $\pm$ 0.0001 & 0.000039 $\pm$ 0.0000112 \\
\hline
& \multicolumn{12}{| c |}{$e^- e^+ \rightarrow ZHH/ZAA$}\\
\hline
$\sigma_{UU}$ & 0.0527 $\pm$ 0.0060 & 0.0278 $\pm$ 0.0030 & 0.0 $\pm$ 0.0 & 0.0 $\pm$ 0.0 & 0.075 $\pm$ 0.011 & 0.062 $\pm$ 0.009 & 0.0152 $\pm$ 0.00200 & 0.0 $\pm$ 0.0 & 0.031 $\pm$ 0.009 & 0.030 $\pm$ 0.008 & 0.02423 $\pm$ 0.0052 & 0.0158 $\pm$ 0.0029 \\
\hline
$\sigma_{LR}$ & 0.0824 $\pm$ 0.0093 & 0.0436 $\pm$ 0.0049 & 0.0 $\pm$ 0.0 & 0.0 $\pm$ 0.0 & 0.117 $\pm$ 0.017 & 0.097 $\pm$ 0.04 & 0.0237 $\pm$ 0.0031 & 0.0 $\pm$ 0.0 & 0.049 $\pm$ 0.015 & 0.047 $\pm$ 0.013 & 0.0379 $\pm$ 0.0082 & 0.0247 $\pm$ 0.0045 \\
\hline
$\sigma_{RL}$ & 0.1284 $\pm$ 0.0146 & 0.0679 $\pm$ 0.0075 & 0.0 $\pm$ 0.0 & 0.0 $\pm$ 0.0 & 0.183 $\pm$ 0.027 & 0.151 $\pm$ 0.022 & 0.0370 $\pm$ 0.0048 & 0.0 $\pm$ 0.0 & 0.077 $\pm$ 0.023 & 0.073 $\pm$ 0.019 & 0.0590 $\pm$ 0.0127 & 0.0386 $\pm$ 0.0071 \\
\hline 
$\sigma_{(0.6, -0.8)}$ & 0.0941 $\pm$ 0.0106 & 0.0498 $\pm$ 0.0054 & 0.0 $\pm$ 0.0 & 0.0 $\pm$ 0.0 & 0.134 $\pm$ 0.019 & 0.110 $\pm$ 0.011 & 0.0271 $\pm$ 0.0035 & 0.0 $\pm$ 0.0 & 0.056 $\pm$ 0.016 & 0.054 $\pm$ 0.014 & 0.0432 $\pm$ 0.0094 & 0.0287 $\pm$ 0.0052 \\
\hline
$\sigma_{(-0.6, 0.8)}$ & 0.0618 $\pm$ 0.0071 & 0.0327 $\pm$ 0.0089 & 0.0 $\pm$ 0.0 & 0.0 $\pm$ 0.0 & 0.088 $\pm$ 0.013 & 0.072 $\pm$ 0.010 & 0.0178 $\pm$ 0.0023 & 0.0 $\pm$ 0.0 & 0.037 $\pm$ 0.011 & 0.035 $\pm$ 0.0095 & 0.0284 $\pm$ 0.0061 & 0.0185 $\pm$ 0.0034 \\
\hline
\end{tabular}
}
\caption{ The cross-section in $fb$ with unpolarized and polarized incoming $e^- e^+$  beams. Where masses of extra Higgs states are varied over a range of  $m_H$ = 150 $GeV$ to $m_H$ = 500 $GeV$. In all cases, distribution of cross-section for both scenarios of with and without polarized beams is the same.}
\label{table:4}
\end{table}


\begin{figure}
\centering     
\subfigure[]{\label{fig:a}\includegraphics[width=53mm]{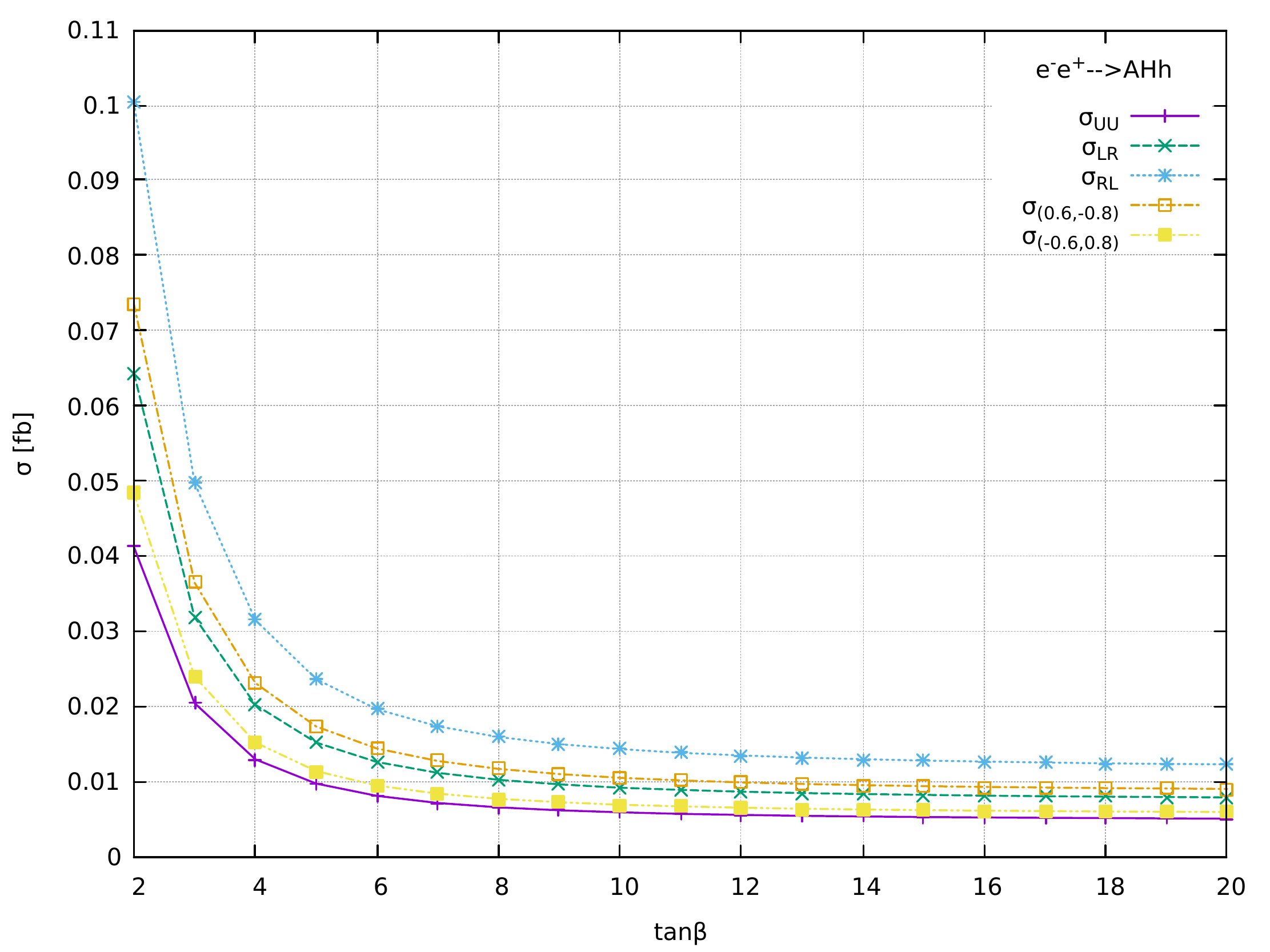}}
\subfigure[]{\label{fig:b}\includegraphics[width=53mm]{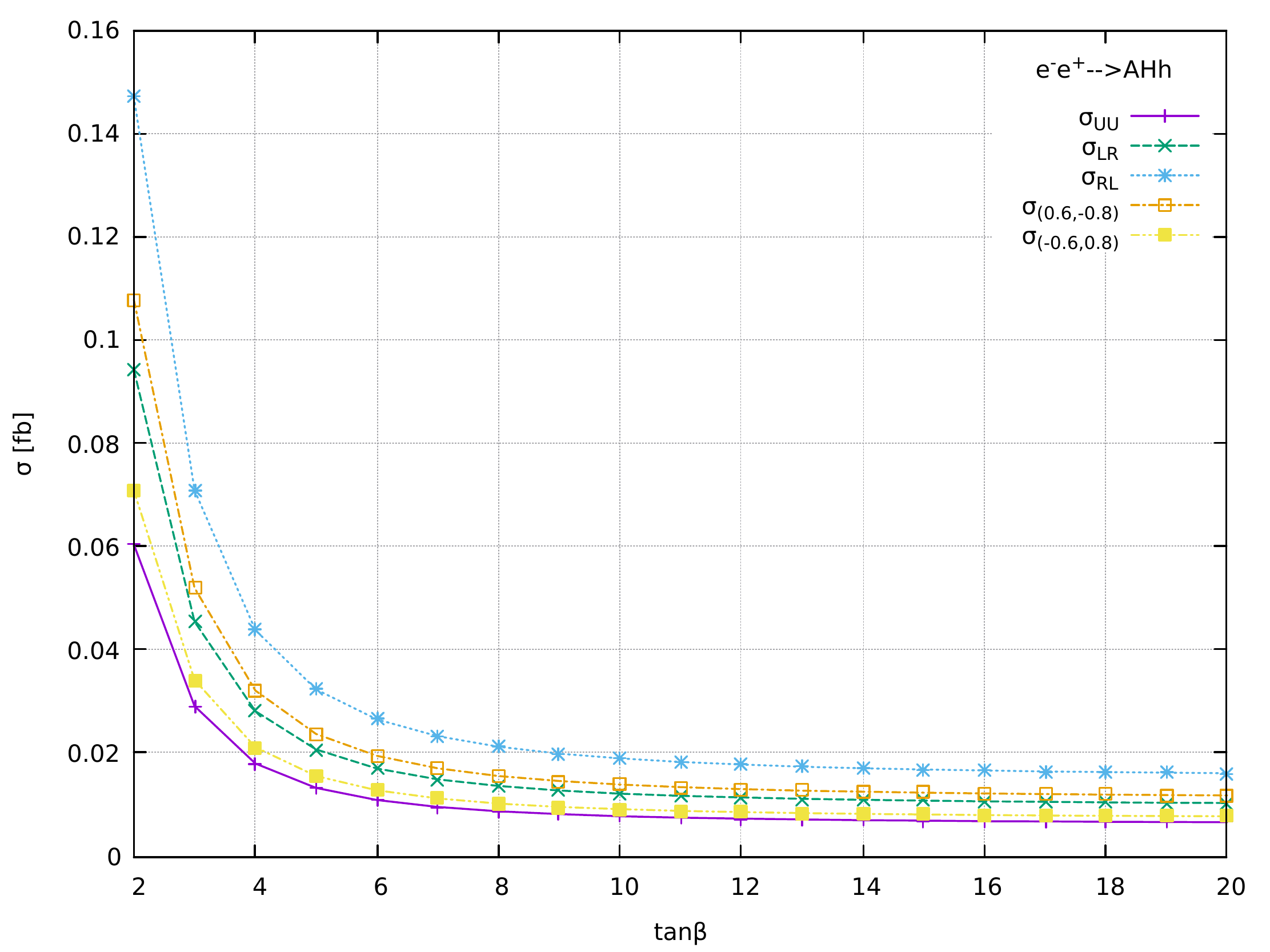}}
\subfigure[]{\label{fig:c}\includegraphics[width=53mm]{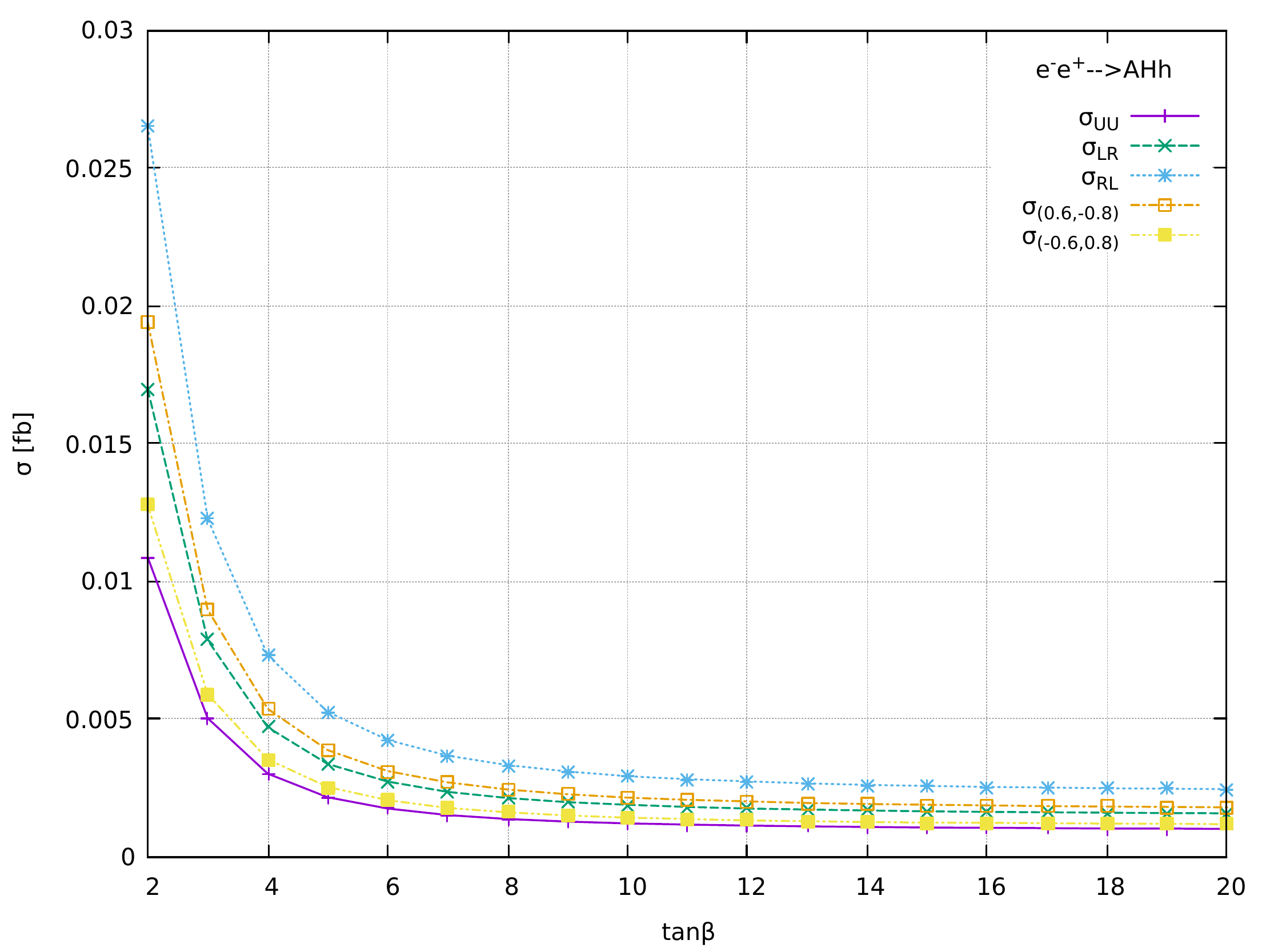}}
\caption{Figure shows dependency of tan$\beta$ on $\sigma(pp \rightarrow AHh)$ at (a) $\sqrt{s}$=0.6 TeV, (b) $\sqrt{s}$=1 TeV and (c) $\sqrt{s}$=3 TeV energies.}
\label{fig:A5}
\end{figure}

\begin{figure}
\centering     
\subfigure[]{\label{fig:a}\includegraphics[width=53mm]{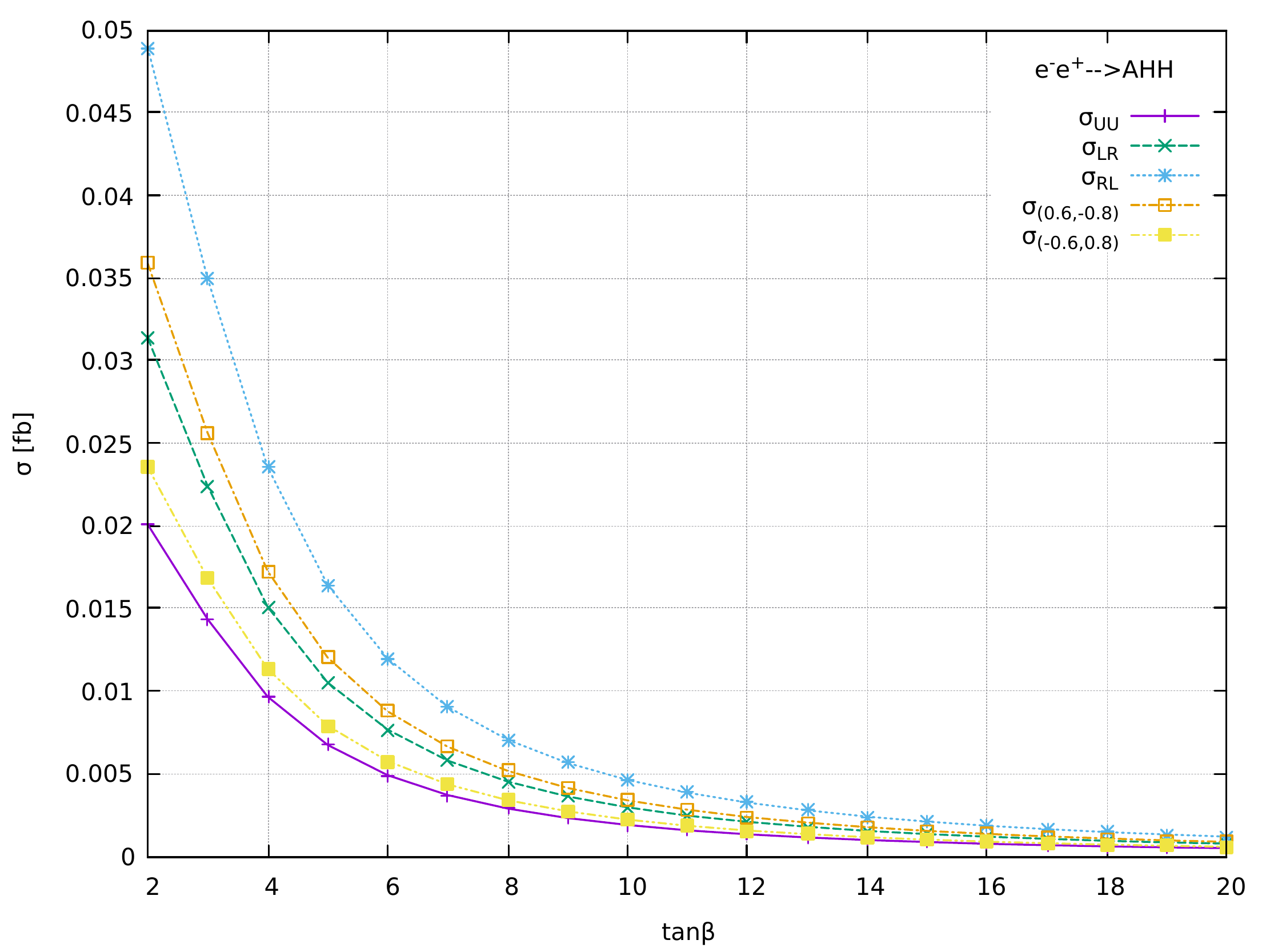}}
\subfigure[]{\label{fig:b}\includegraphics[width=53mm]{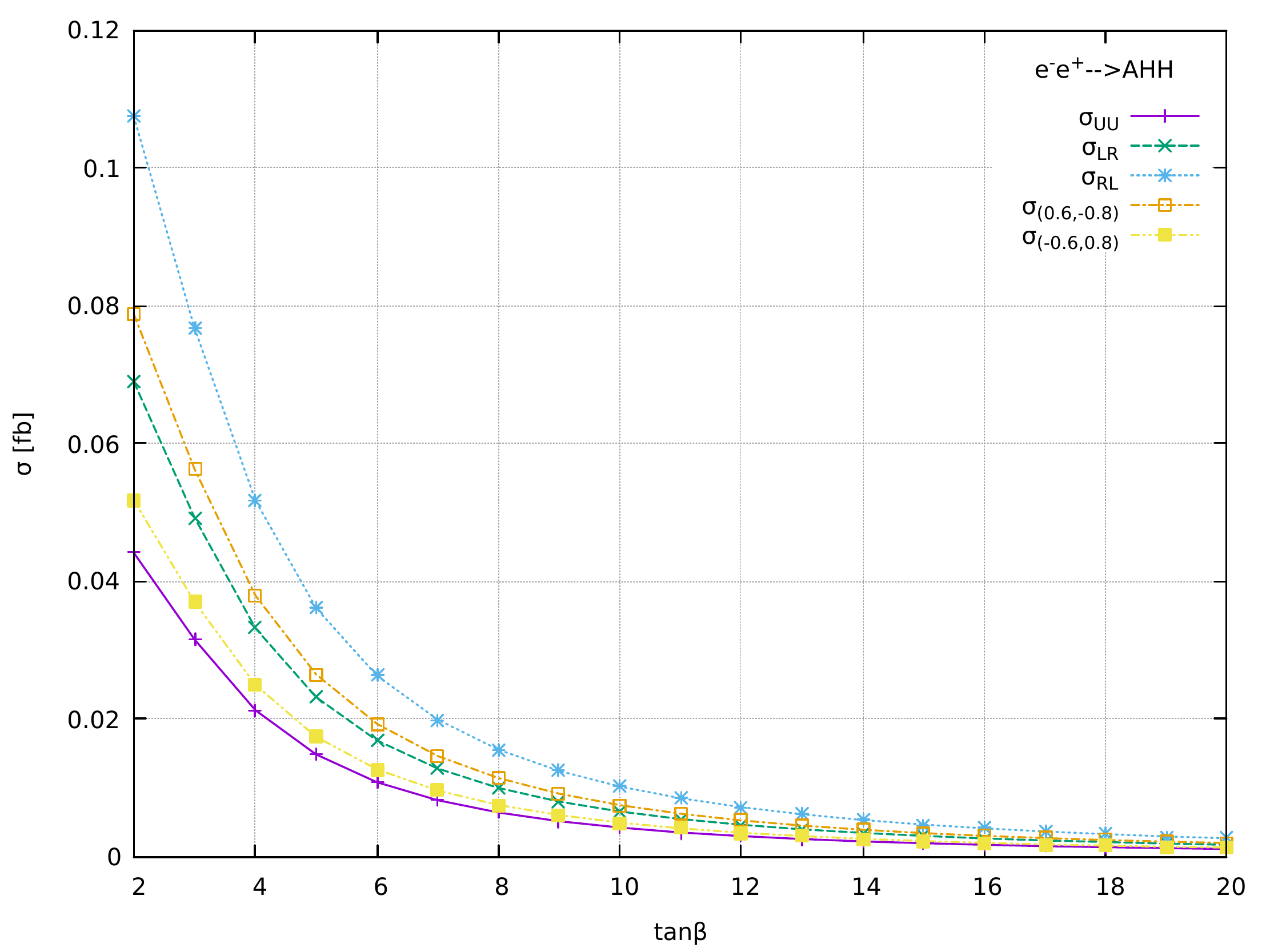}}
\subfigure[]{\label{fig:c}\includegraphics[width=53mm]{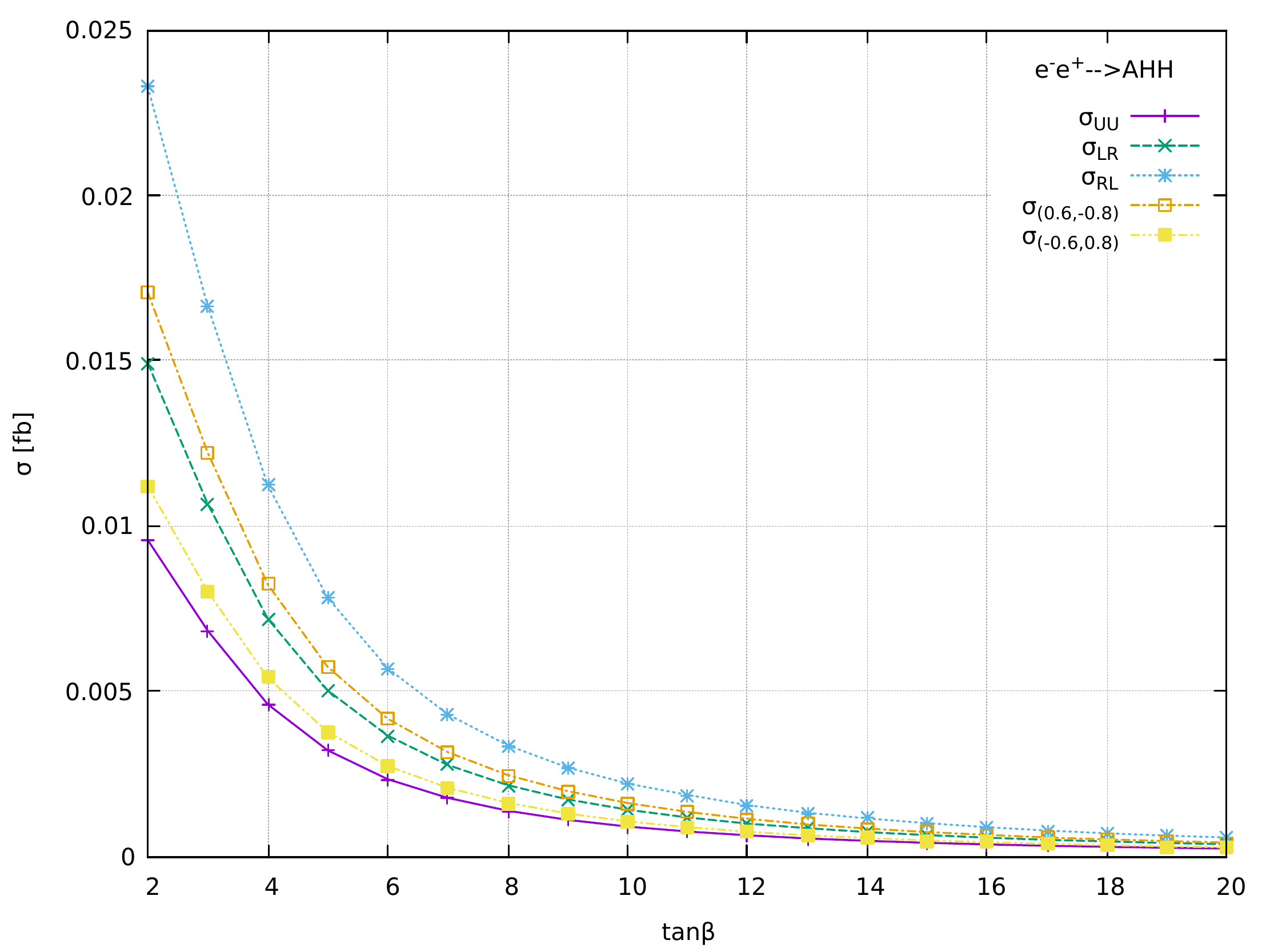}}
\caption{In Figures the production cross sections $\sigma(pp \rightarrow AHH)$ are shown as a function of tan$\beta$ at (a) 0.6 TeV, (b) 1 TeV and (c) 3 TeV energies.}
\label{fig:B5}
\end{figure}

\begin{figure}
\centering     
\subfigure[]{\label{fig:a}\includegraphics[width=53mm]{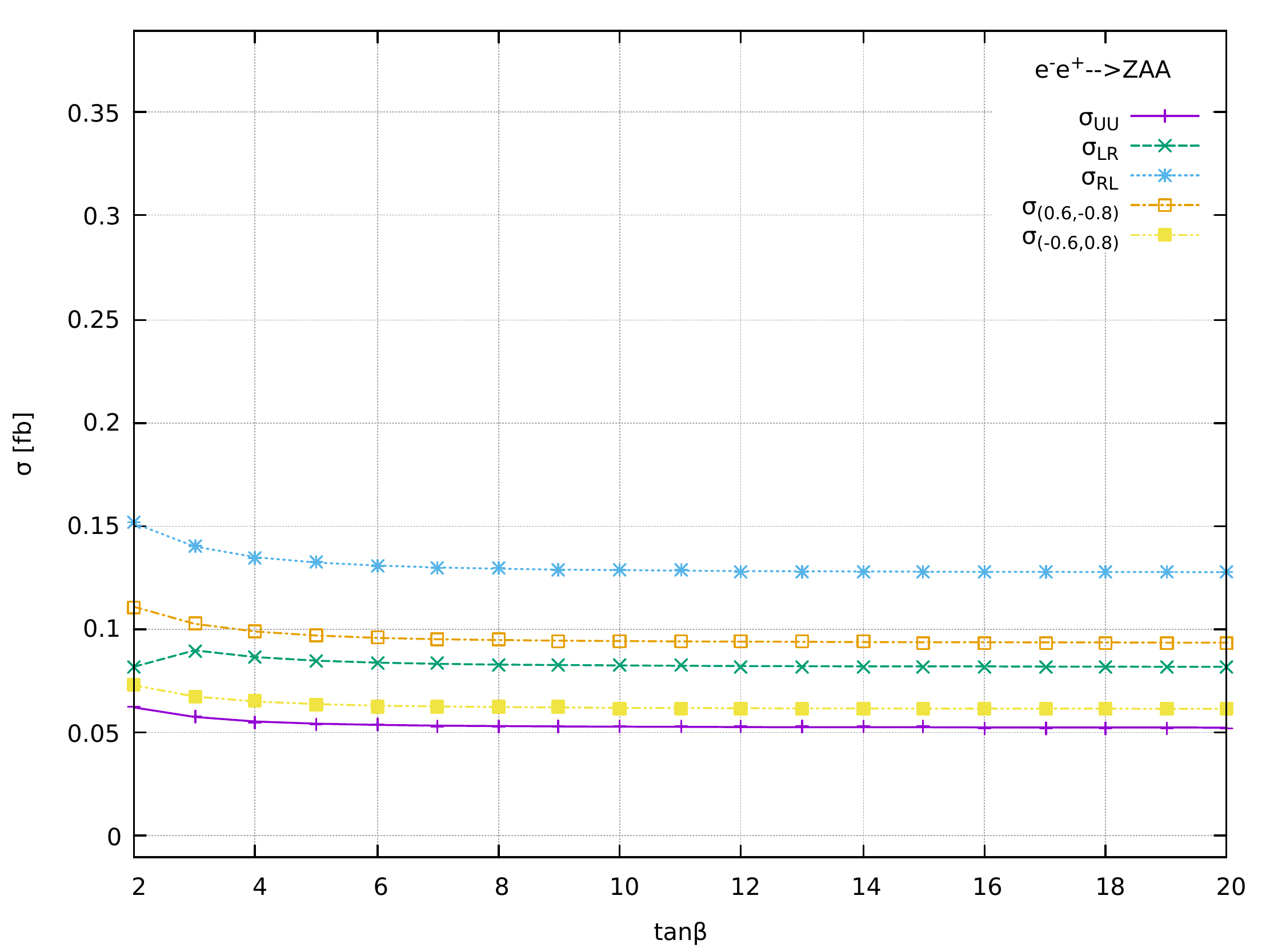}}
\subfigure[]{\label{fig:b}\includegraphics[width=53mm]{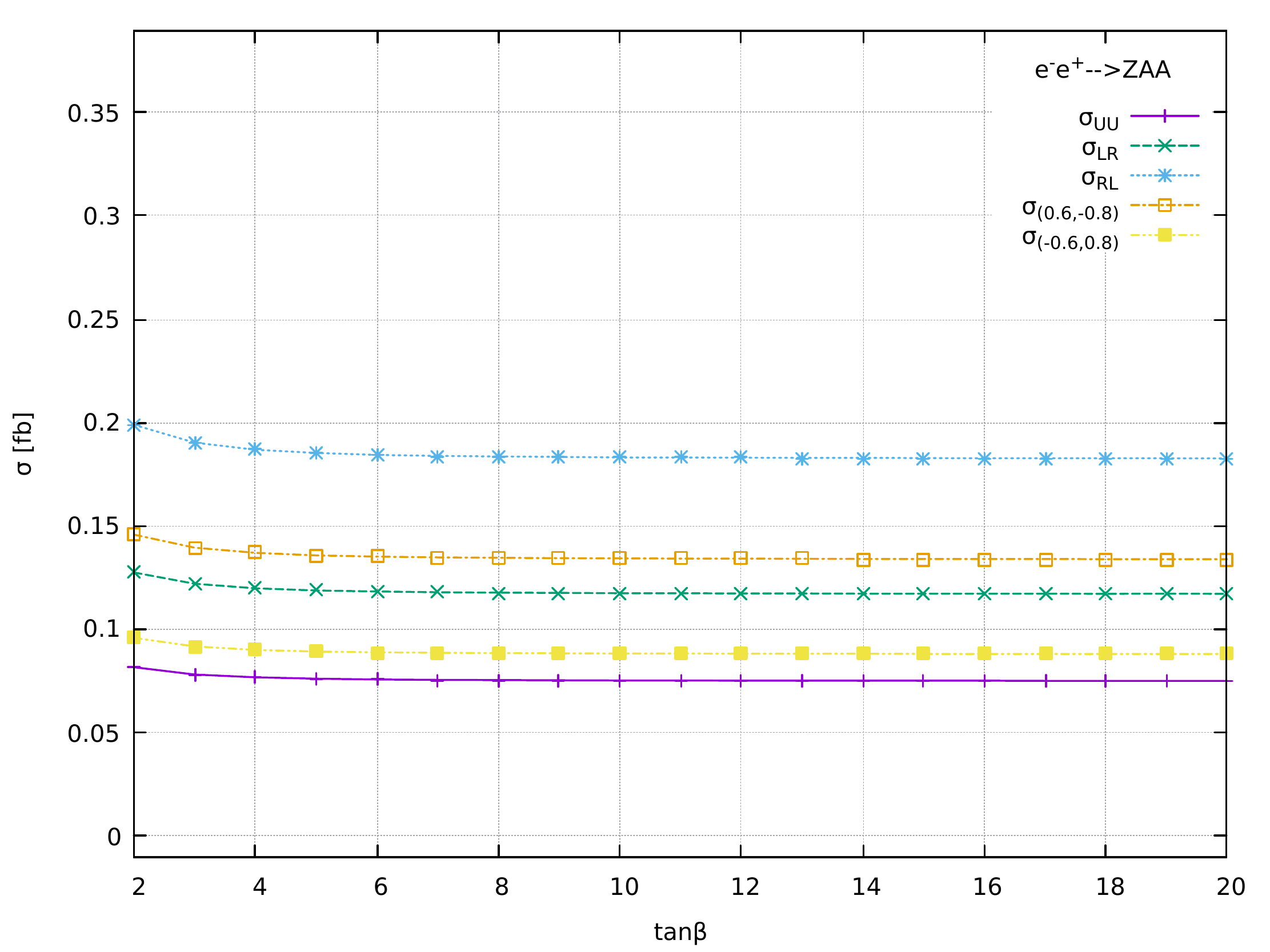}}
\subfigure[]{\label{fig:c}\includegraphics[width=53mm]{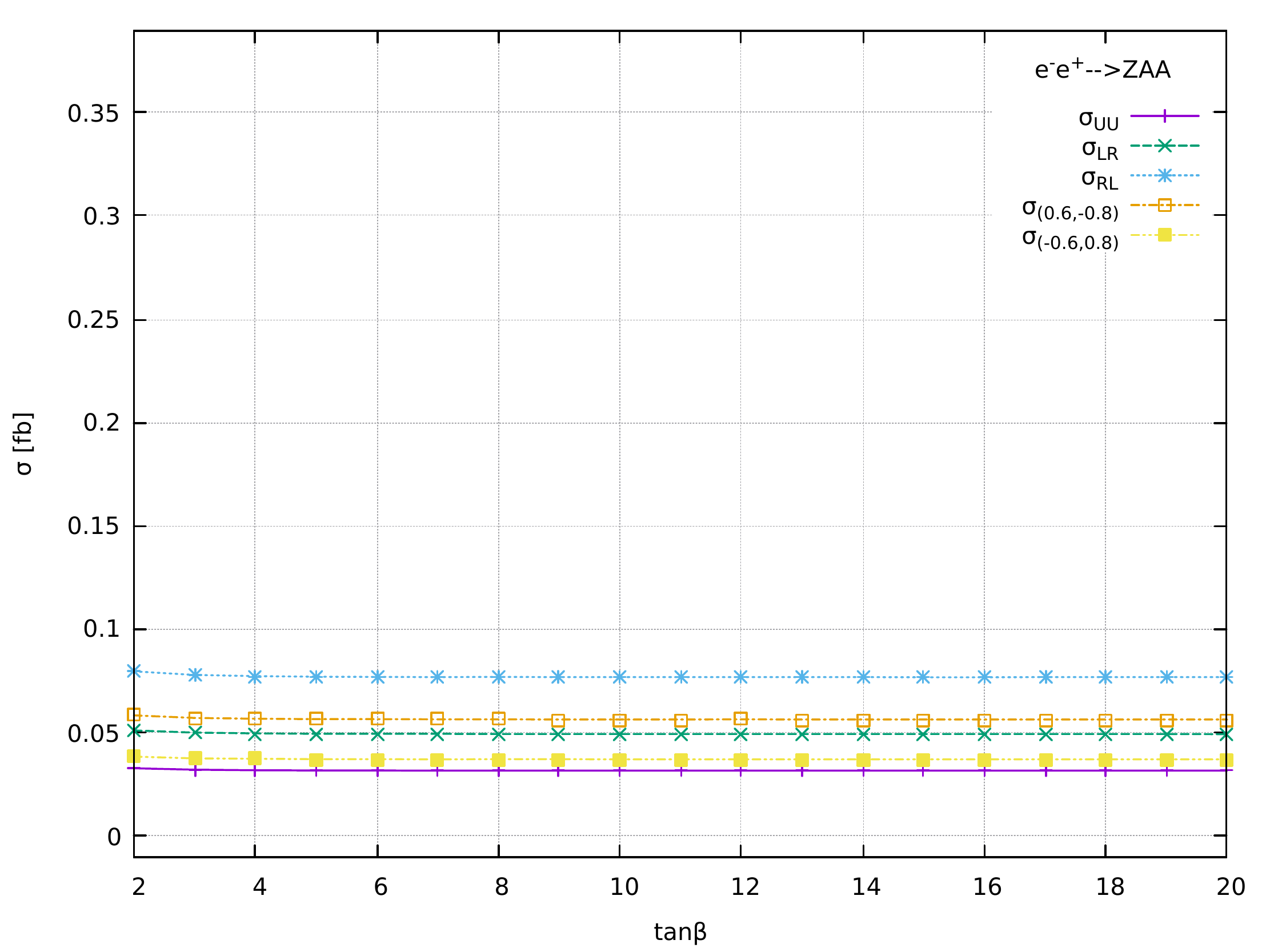}}
\caption{Figure (a) is plotted to discuss the bechaviour of $\sigma(pp \rightarrow ZAA)$ at $\sqrt{s}$=0.6 TeV, in whereas (b) and (c) are plotted same variables at $\sqrt{s}$=1 TeV and $\sqrt{s}$=3 TeV respectively}
\label{fig:C5}
\end{figure}

\begin{figure}
\centering     
\subfigure[]{\label{fig:a}\includegraphics[width=53mm]{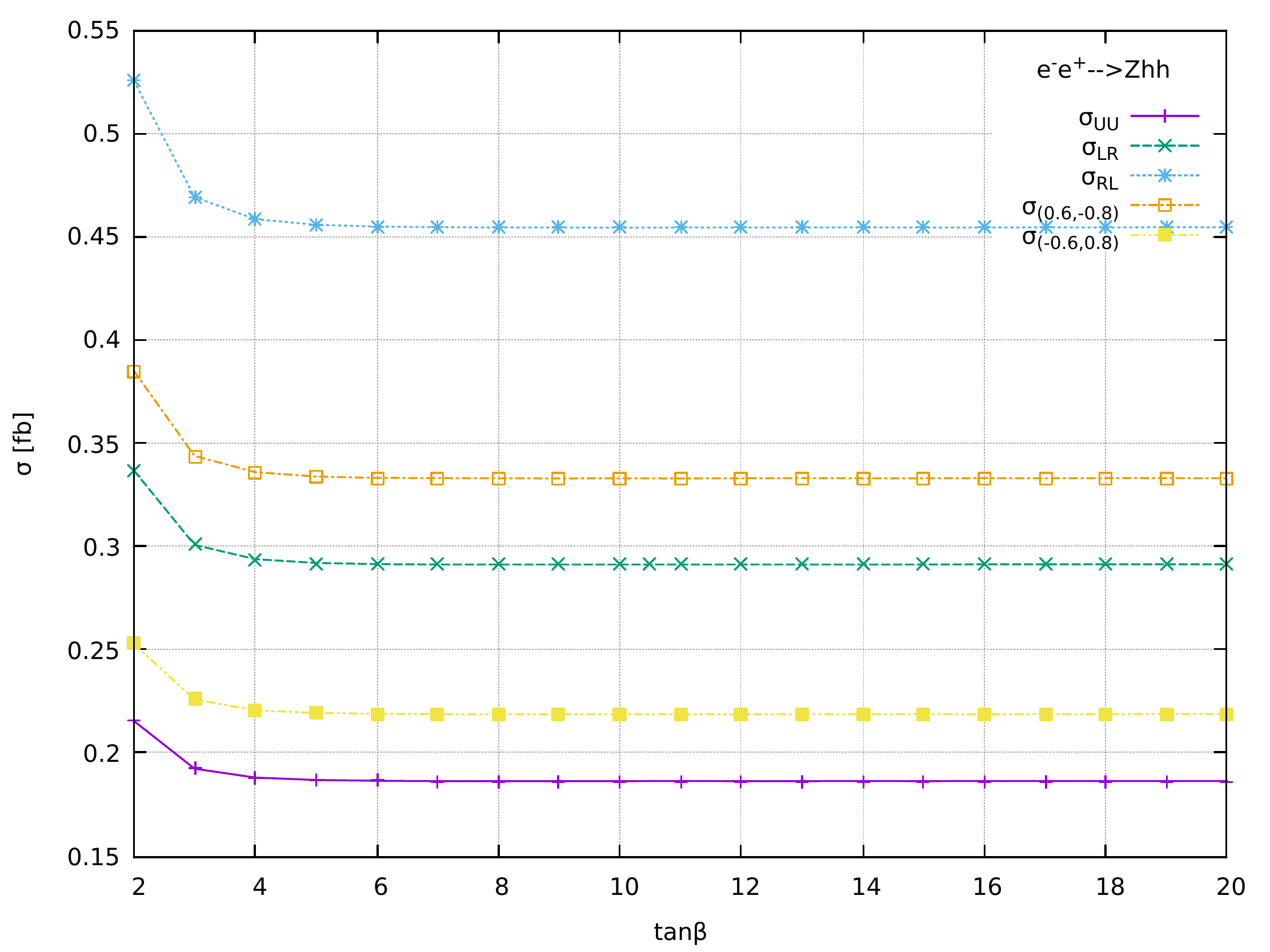}}
\subfigure[]{\label{fig:b}\includegraphics[width=53mm]{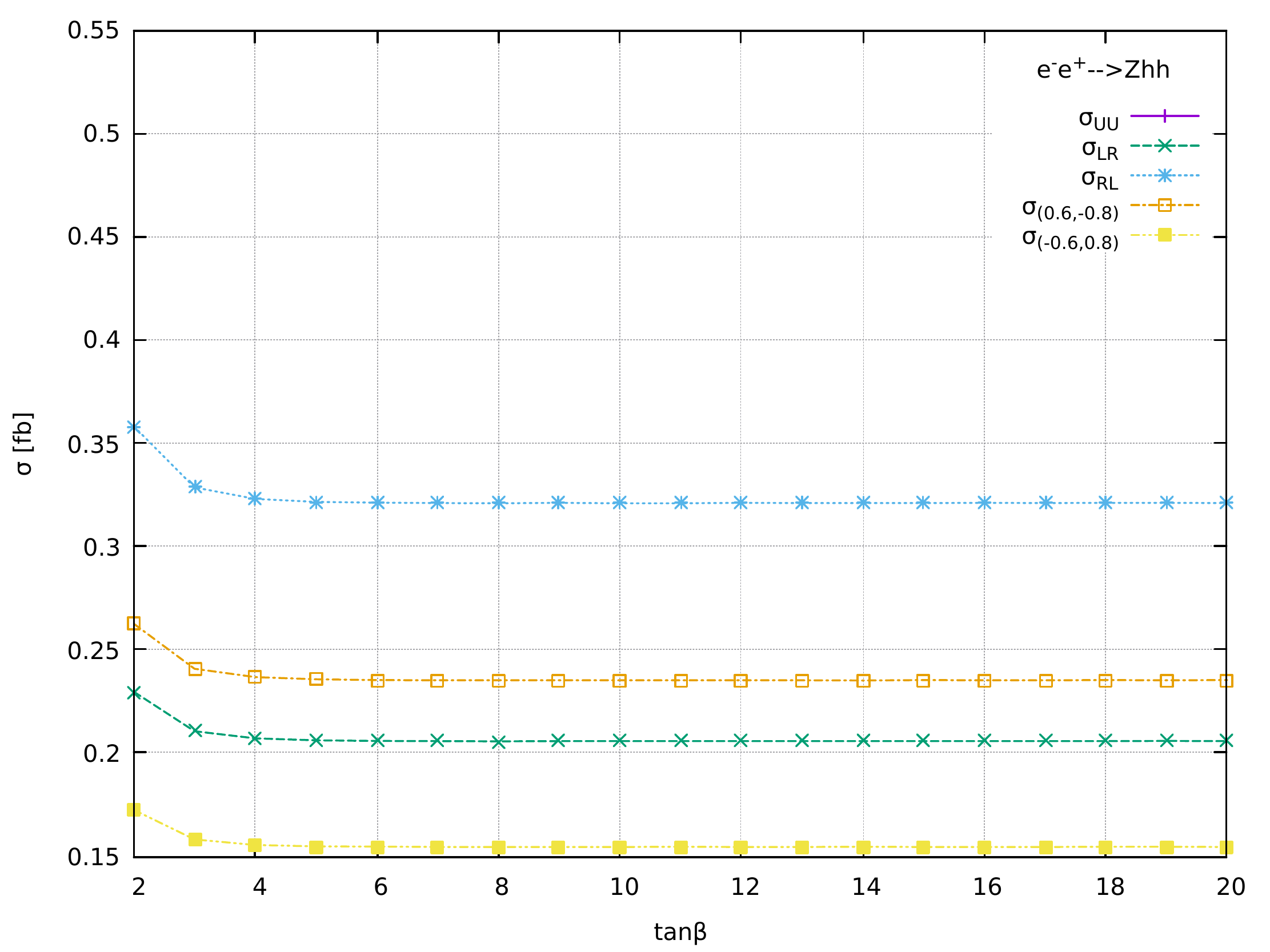}}
\subfigure[]{\label{fig:c}\includegraphics[width=53mm]{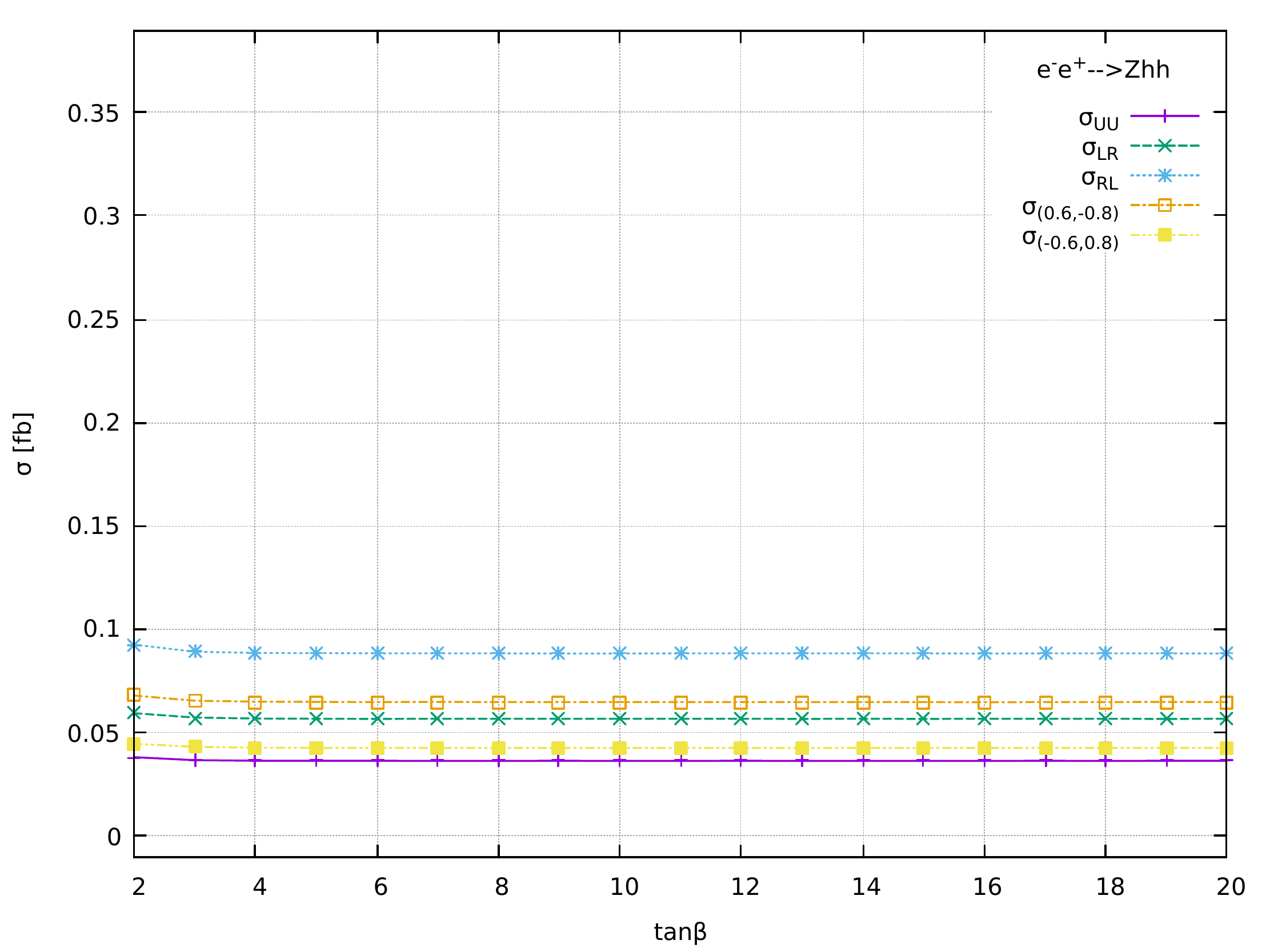}}
\caption{Figure shows dependency of tan$\beta$ on $\sigma(pp \rightarrow Zhh)$ at (a) $\sqrt{s}$=0.6 TeV, (b) $\sqrt{s}$=1 TeV and (c) $\sqrt{s}$=3 TeV energies.}
\label{fig:E5}
\end{figure}

\begin{figure}
\centering     
\subfigure[]{\label{fig:a}\includegraphics[width=53mm]{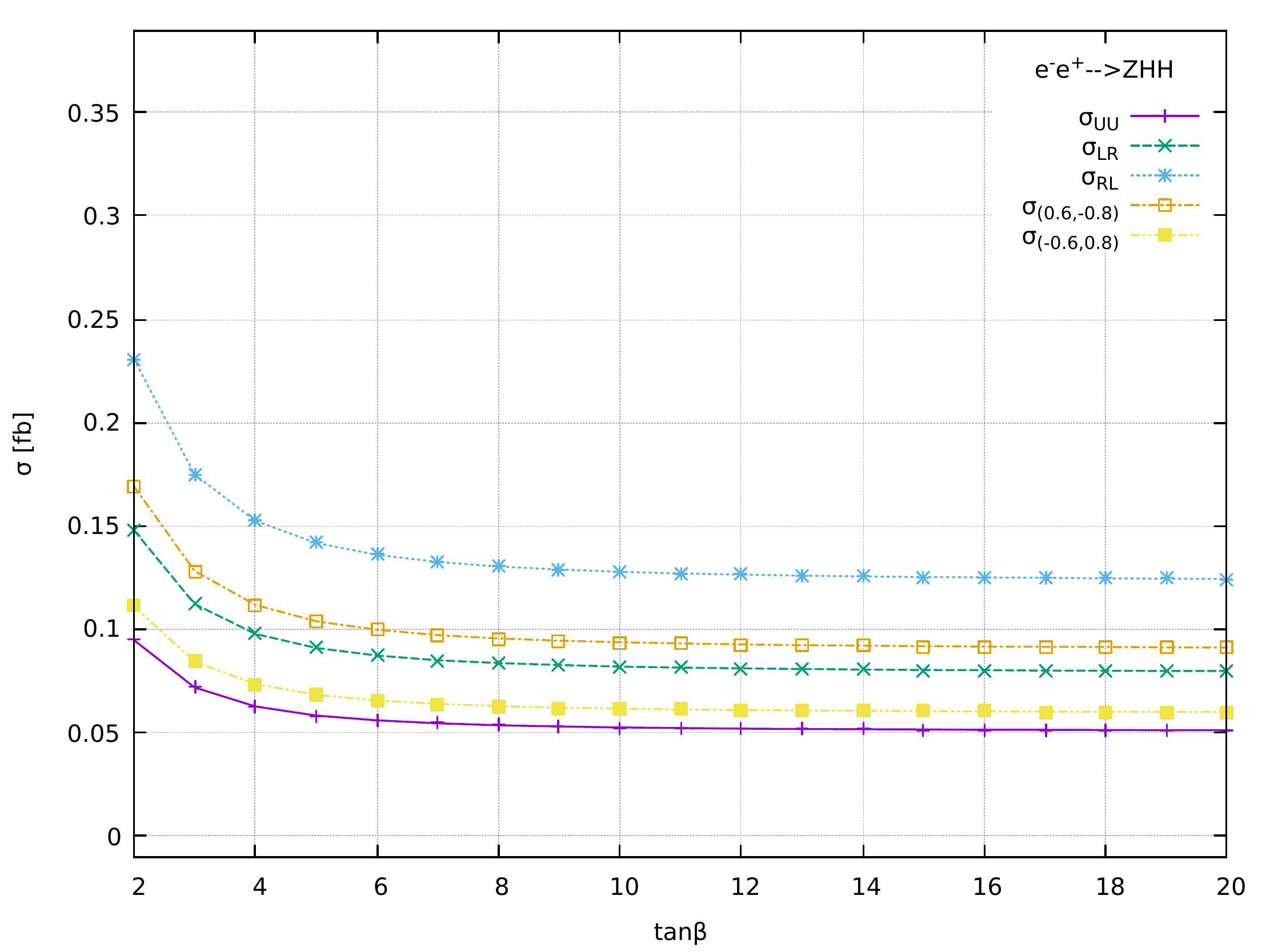}}
\subfigure[]{\label{fig:b}\includegraphics[width=53mm]{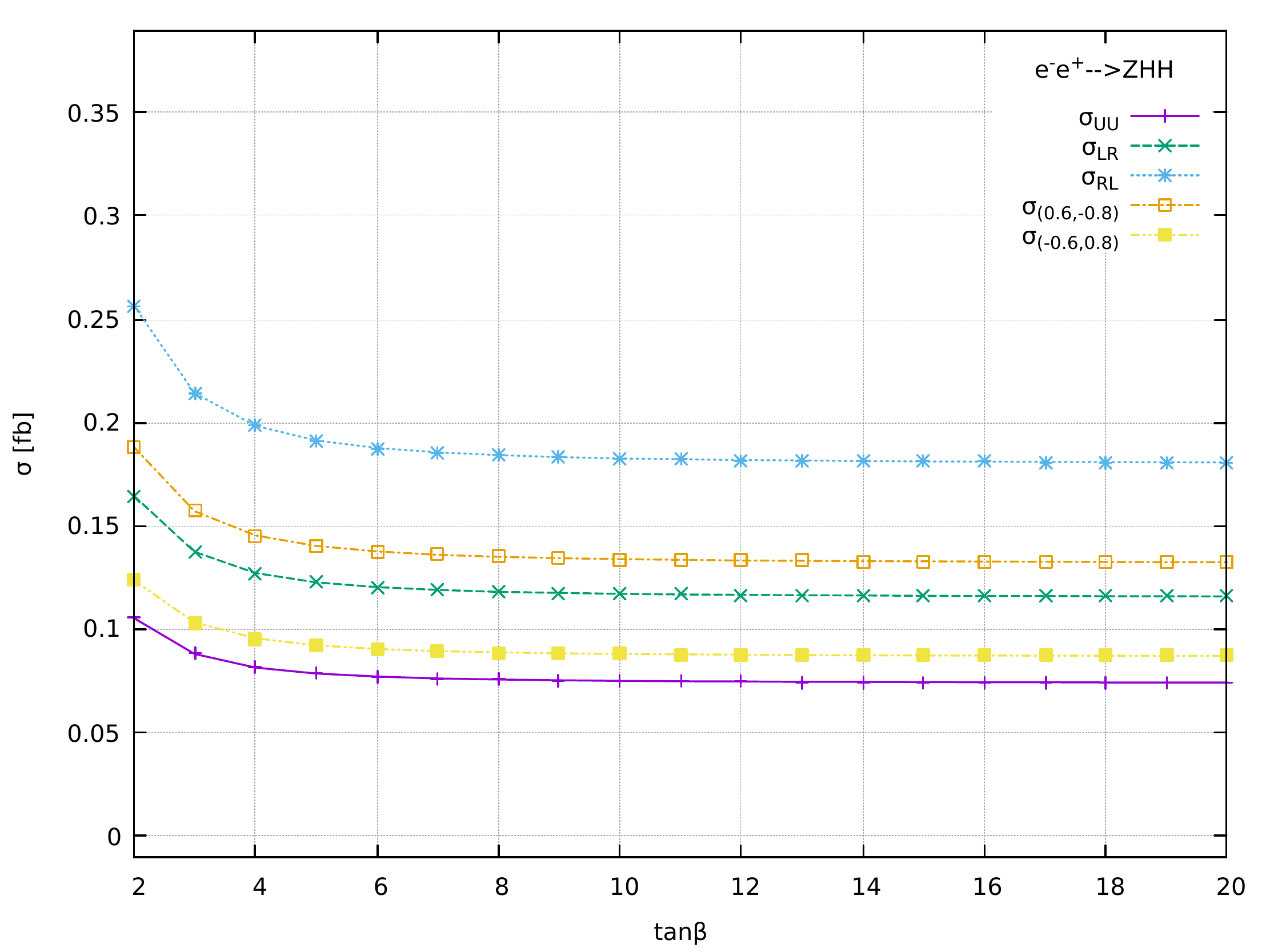}}
\subfigure[]{\label{fig:c}\includegraphics[width=53mm]{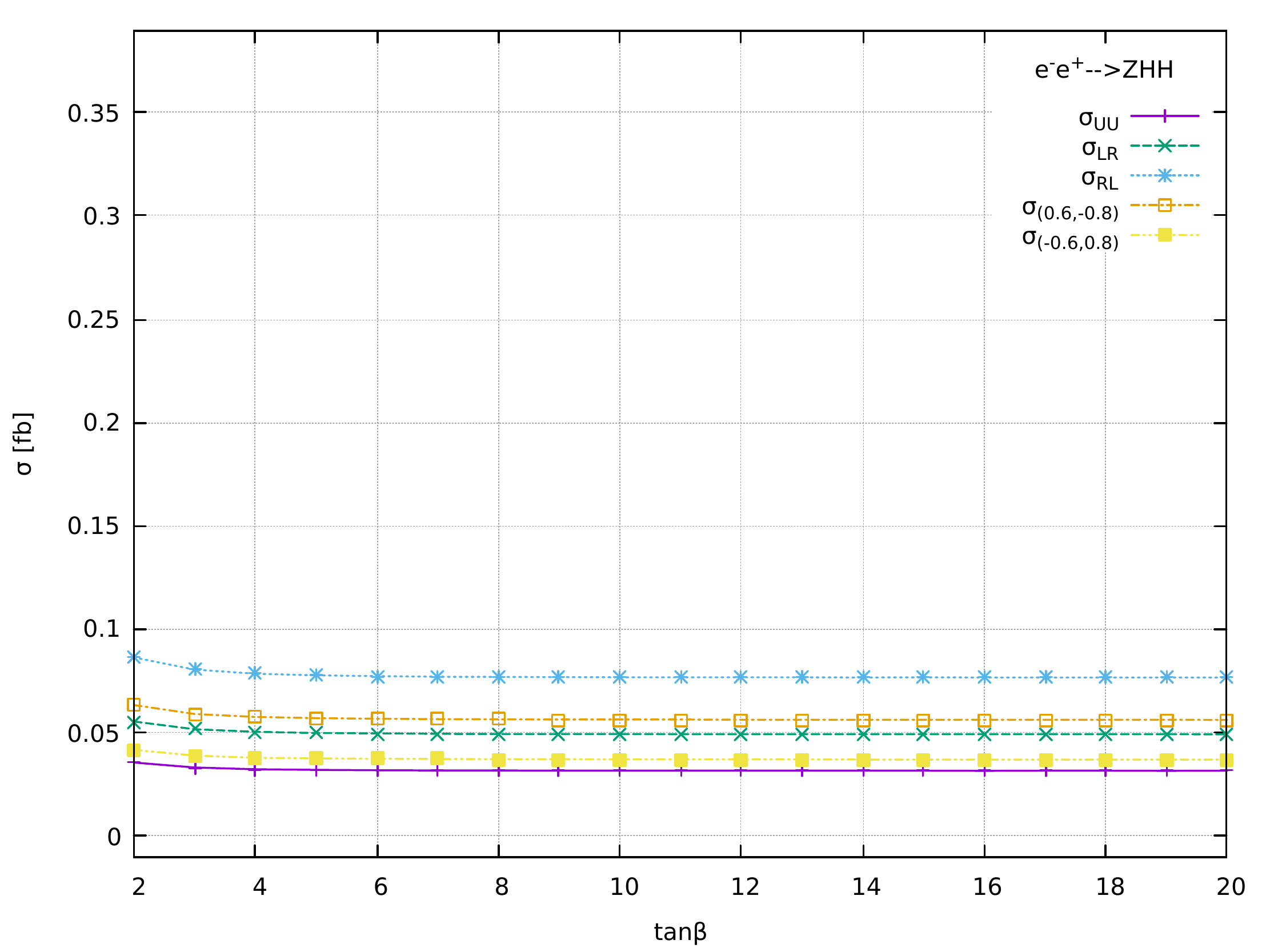}}
\caption{In Figures the production cross sections $\sigma(pp \rightarrow ZHH)$ are shown as a function of tan$\beta$ at (a) 0.6 TeV, (b) 1 TeV and (c) 3 TeV energies.}
\label{fig:E5}
\end{figure}

\begin{table}[H]
\begin{center}
\resizebox{17cm}{!}{
 \begin{tabular}{|c|c|c|c|c|c|c|c|c|c|c|c|c|c|c|c|c|}
\hline
  & \multicolumn{4}{| c |}{0.6 TeV}   & \multicolumn{4}{| c |}{1 TeV} & \multicolumn{4}{| c |}{3 TeV} \\
\hline
 &  $\tan{\beta}$=5 & $\tan{\beta}$=10 &  $\tan{\beta}$=15 &  $\tan{\beta}$=20 &  $\tan{\beta}$=5 & $\tan{\beta}$=10 &  $\tan{\beta}$=15 &  $\tan{\beta}$=20 &  $\tan{\beta}$=5 & $\tan{\beta}$=10 &  $\tan{\beta}$=15 &  $\tan{\beta}$=20 \\
\hline
& \multicolumn{12}{| c |}{$e^{-} e^{+} \rightarrow Zhh$}\\
\hline
 $\sigma_{UU}$ & 0.1869 & 0.1863 & 0.1864 & 0.186 & 0.1319 & 0.1316 & 0.1317 & 0.1317 & 0.036253 & 0.0361733 & 0.03616 & 0.0362 \\
 \hline
 $\sigma_{LR}$ & 029203 & 0.29120 & 0.29126 & 0.29130 & 0.20615 & 0.20572 & 0.20573 & 0.20573 & 0.056621 & 0.056543 & 0.056539 & 0.056557 \\
 \hline
 $\sigma_{RL}$ & 0.45565 & 0.45426 & 0.45435 & 0.45442 & 0.31259 & 0.32089 & 0.32096 & 0.32098 & 0.088306 & 0.088241 & 0.088241 & 0.088193 \\
\hline
$\sigma_{(0.6, -0.8)}$ & 0.33387 & 0.33297 & 0.33301 & 0.33299 & 0.23570 & 0.23521 & 0.23530 & 0.23525 & 0.064696 & 0.064651 & 0.064676 & 0.064674 \\
\hline
$\sigma_{(-0.6, 0.8)}$ & 0.21934 & 0.21877 & 0.21878 & 0.21880 & 0.15479 & 0.15452 & 0.15454 & 0.15455 & 0.042510 & 0.042491 & 0.042457 & 0.042514 \\
\hline
& \multicolumn{12}{| c |}{$e^{-} e^{+} \rightarrow AHh$}\\
\hline
 $\sigma_{UU}$ & 0.017277 & 0.01045 & 0.00093411 & 0.0089632 & 0.013257 & 0.0077706 & 0.0068967 & 0.0065998 & 0.0021479 & 0.001047 & 0.001047 & 0.00099589 \\
\hline
 $\sigma_{LR}$ & 0.015155 & 0.0091414 & 0.0081716 & 0.0078409 & 0.020702 & 0.01214 & 0.010770 & 0.010317 & 0.0033532 & 0.0018687 & 0.0016357 & 0.0015611 \\
\hline
 $\sigma_{RL}$ & 0.023581 & 0.014264 & 0.012748 & 0.012232 & 0.032320 & 0.018947 & 0.016800 & 0.016090 & 0.0052484 & 0.0029070 & 0.0025522 & 0.0024360 \\
\hline
$\sigma_{(0.6, -0.8)}$ & 0.017277 & 0.010454 & 0.0093411 & 0.0089632 & 0.023686 & 0.013889 & 0.012320 & 0.011787 & 0.0038544 & 0.0021296 & 0.0018715 & 0.0017796 \\
\hline 
$\sigma_{(-0.6, 0.8)}$ & 0.011354 & 0.0068721 & 0.0061365 & 0.0058895 & 0.015560 & 0.0091174 & 0.0080898 & 0.0077460 & 0.0025196 & 0.0014044 & 0.0012263 & 0.0011680 \\
\hline
& \multicolumn{12}{| c |}{$e^{-} e^{+} \rightarrow AHH$}\\
\hline
$\sigma_{UU}$ & 0.0067367 & 0.001894 & 0.0008613 & 0.00048807 & 0.014793 & 0.0064978 & 0.0018905 & 0.001071 & 0.00321 & 0.00090088 & 0.00041045 & 0.00023214 \\
 \hline
 $\sigma_{LR}$ & 0.010523 & 0.0029595 & 0.0013454 & 0.00076264 & 0.023111 & 0.0064978 & 0.0029554 & 0.0016753 & 0.0050175 & 0.0014085 & 0.00063930 & 0.00036364 \\
 \hline
 $\sigma_{RL}$ & 0.016417 & 0.0046182 & 0.0020992 & 0.001897 & 0.036058 & 0.010142 & 0.0046089 & 0.0026143 & 0.0078280 & 0.0022009 & 0.0010019 & 0.00056732 \\
\hline
$\sigma_{(0.6, -0.8)}$ & 0.012032 & 0.0033842 & 0.0015387 & 0.00087216 & 0.026419 & 0.0074374 & 0.0033764 & 0.0019149 & 0.0057319 & 0.0016118 & 0.00073268 & 0.00041471 \\
\hline 
$\sigma_{(-0.6, 0.8)}$ & 0.0079058 & 0.0022240 & 0.0010107 & 0.00057310 & 0.017364 & 0.0048834 & 0.0022192 & 0.0012581 & 0.0037654 & 0.0010560 & 0.00048088 & 0.00027258 \\
\hline
& \multicolumn{12}{| c |}{$e^{-} e^{+} \rightarrow ZHH$}\\
\hline
 $\sigma_{UU}$ & 0.058131 & 0.052392 & 0.051358 & 0.050993 & 0.078495 & 0.074928 & 0.07438 & 0.07412 & 0.031811 & 0.03148 & 0.031411 & 0.031407 \\
\hline 
 $\sigma_{LR}$ & 0.090831 & 0.0818630 & 0.080237 & 0.079662 & 0.12264 & 0.11703 & 0.11611 & 0.11579 & 0.049703 & 0.049168 & 0.049067 & 0.049034 \\
\hline
 $\sigma_{RL}$ & 0.14170 & 0.12771 & 0.12517 & 0.1247 & 0.19129 & 0.18261 & 0.18117 & 0.18065 & 0.077557 & 0.076655 & 0.076611 & 0.076518 \\
\hline
$\sigma_{(0.6, -0.8)}$ & 0.10384 & 0.093580 & 0.091721 & 0.091077 & 0.14022 & 0.13385 & 0.13275 & 0.13237 & 0.056871 & 0.056211 & 0.056085 & 0.056052 \\
\hline
$\sigma_{(-0.6, 0.8)}$ & 0.068235 & 0.061480 & 0.060274 & 0.059846 & 0.092134 & 0.087936 & 0.087239 & 0.086994 & 0.037366 & 0.036951 & 0.036839 & 0.036841 \\
\hline
& \multicolumn{12}{| c |}{$e^{-} e^{+} \rightarrow ZAA$}\\
\hline
 $\sigma_{UU}$ & 0.054229 & 0.05272 & 0.052419 & 0.052313 & 0.075971 & 0.075097 & 0.074993 & 0.074924 & 0.031578 & 0.031489 & 0.031469 & 0.031467 \\
\hline
 $\sigma_{LR}$ & 0.084719 & 0.08235 & 0.081905 & 0.081737 & 0.11868 & 0.11735 & 0.11712 & 0.11702 & 0.049323 & 0.049201 & 0.049158 & 0.049131 \\
\hline
 $\sigma_{RL}$ & 0.13219 & 0.12851 & 0.12777 & 0.12751 & 0.18521 & 0.18306 & 0.18277 & 0.18261 & 0.076973 & 0.076717 & 0.76681 & 0.076691 \\
\hline
$\sigma_{(0.6, -0.8)}$ & 0.096858 & 0.094178 & 0.093642 & 0.093460 & 0.13569 & 0.13418 & 0.13389 & 0.13384 & 0.056379 & 0.056242 & 0.056228 & 0.056194 \\
\hline
$\sigma_{(-0.6, 0.8)}$ & 0.063644 & 0.061869 & 0.061510 & 0.061416 & 0.089156 & 0.088173 & 0.087995 & 0.087925 & 0.037060 & 0.036940 & 0.036912 & 0.036923 \\
\hline
\end{tabular}%
}
\caption{ A complete scan of most relevant parameters of 5 different production processes at Higgs mass $m_\phi$ =150 GeV is performed }
\label{table:3a}
\end{center}
\end{table}

\subsubsection{$e^- e^+ \rightarrow ZHH / ZAA$}

\begin{figure}
\centering     
\subfigure[]{\label{fig:a}\includegraphics[width=60mm]{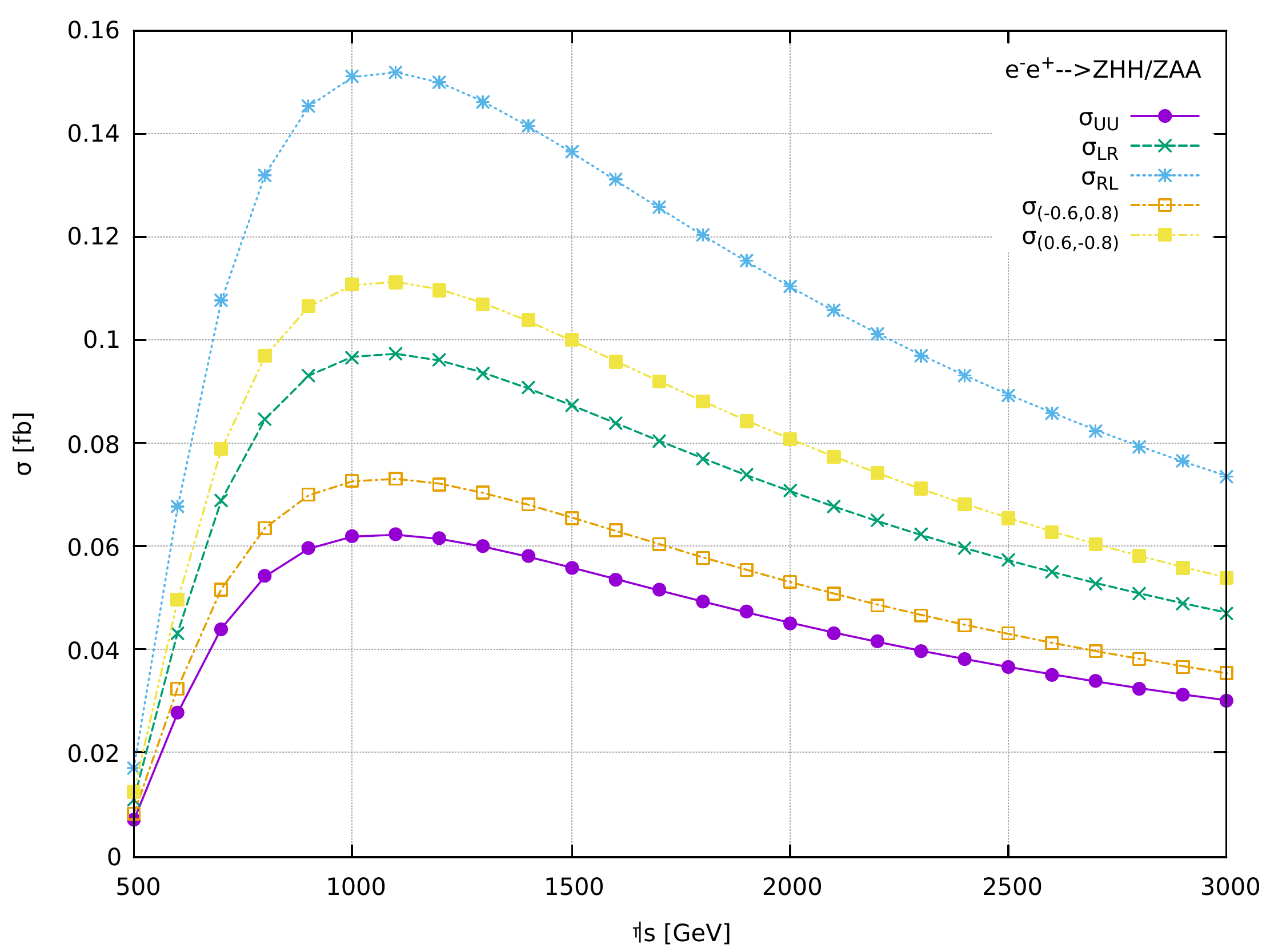}}
\subfigure[]{\label{fig:b}\includegraphics[width=60mm]{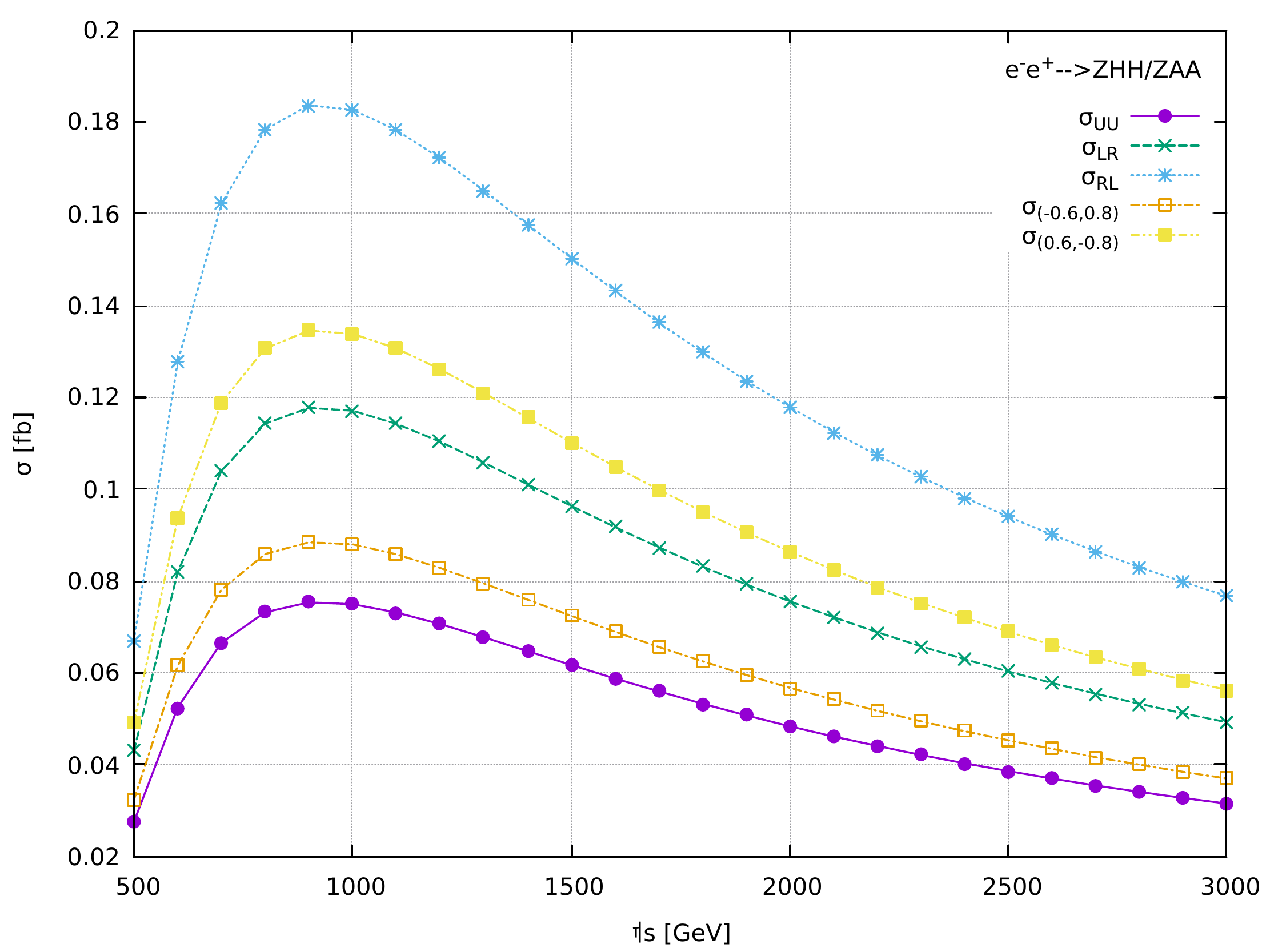}}
\subfigure[]{\label{fig:a}\includegraphics[width=60mm]{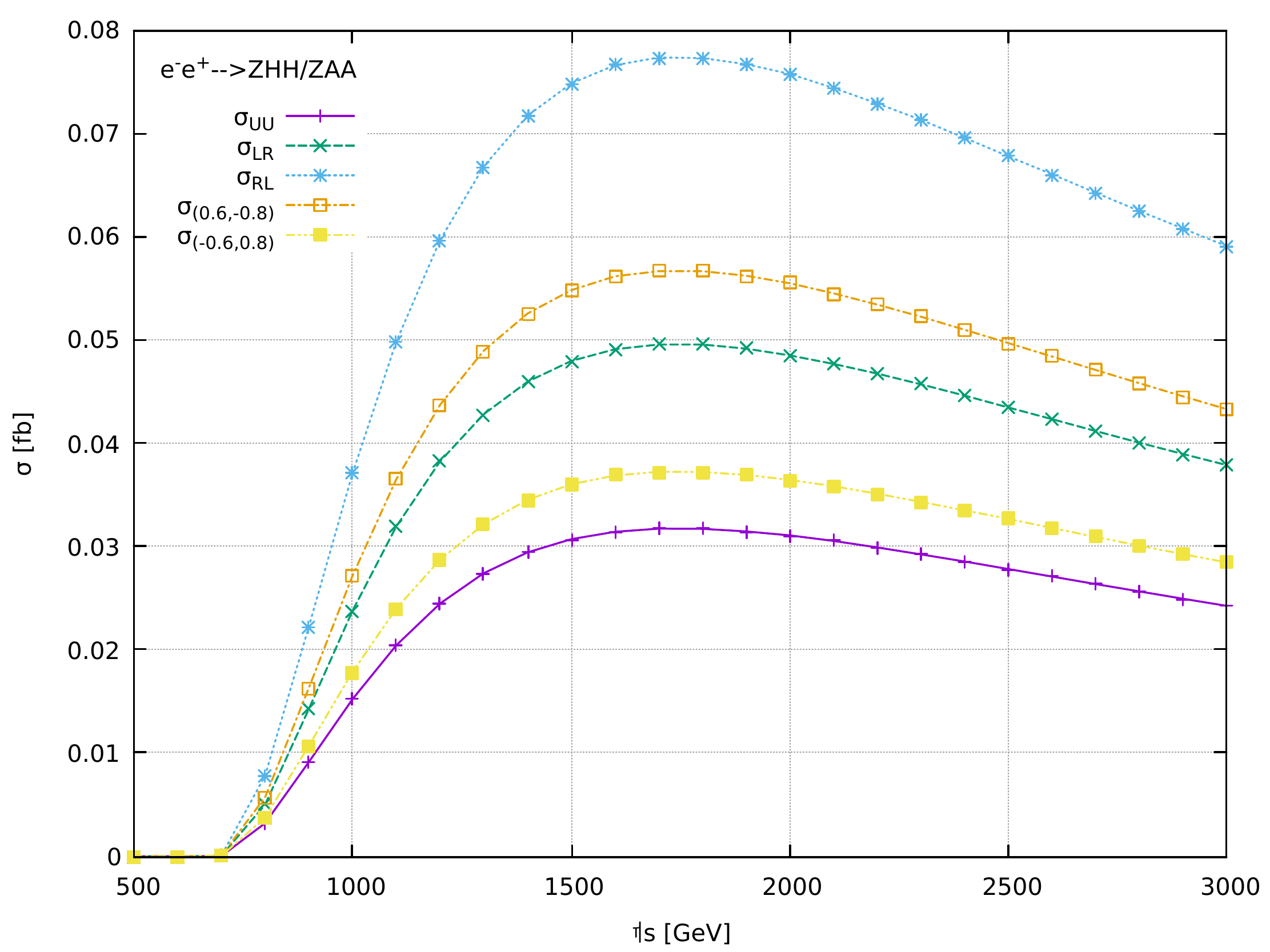}}
\subfigure[]{\label{fig:b}\includegraphics[width=60mm]{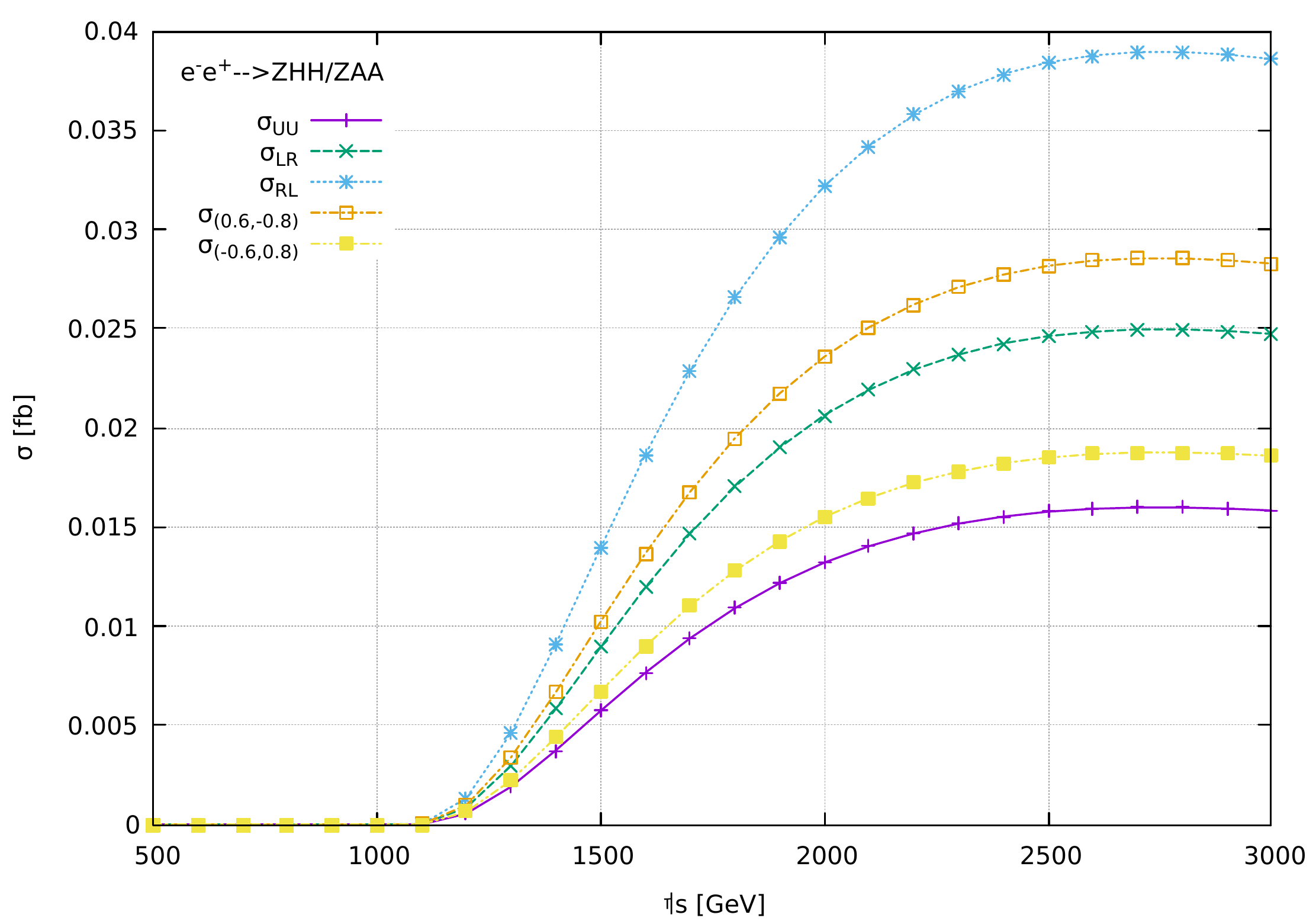}}
\caption{Total cross-section $\sigma$ ($fb$) as a function of $\sqrt{s}$ ($GeV$) at 2HDM. Higgs mass value $m_H$ = 175 $GeV$, $\tan{\beta}$ = 10 and $s_{\beta \alpha}$ = 1 are used (left). The Higgs masses $m_H$ = 150 $GeV$, $\tan{\beta}$ = 10 and $s_{\beta \alpha}$ = 1 are used (right)}
\label{fig:6_5.5}
\end{figure}

\begin{figure}[h]
    \centering
    \includegraphics[scale=0.50]{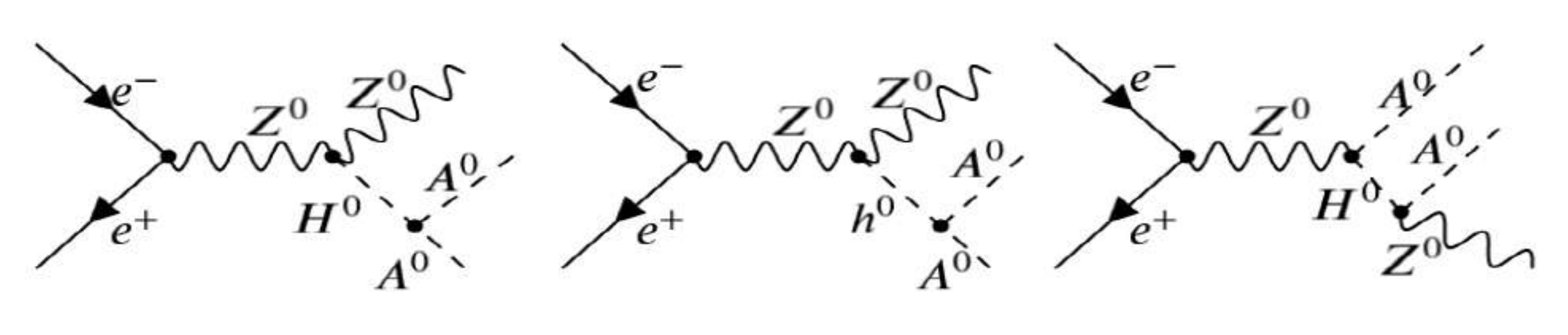}
    \caption{Feynman diagrams contributing to the scattering process   $e^- e^+ \rightarrow \text{ZAA}$.}
    \label{fig:7_5.6}
\end{figure}

\begin{figure}[h]
    \centering
    \includegraphics[scale=0.50]{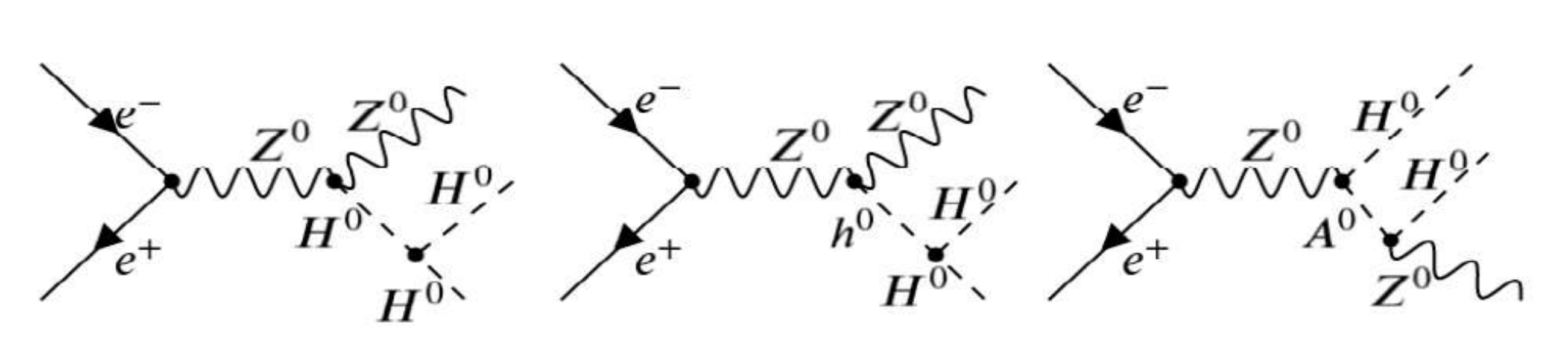}
    \caption{Feynman diagrams contributing to the scattering process   $e^- e^+ \rightarrow \text{ZHH}$ at ILC.}
    \label{fig:8_5.7}
\end{figure}

Consider the scattering process $e^- e^+ \rightarrow ZHH / ZAA$ for determination of cross-section distribution. Since according to Equations ~\ref{eq:4.5b} and ~\ref{eq:4.6b}, the couplings $g_{h^0 A^0 A^0}$  and  $g_{h^0 H^0 H^0}$ of both scattering processes are same, and in terms of parameter space, $m_H$ and $m_A$  are also identical, hence these two scattering process $e^- e^+ \rightarrow ZHH$ and $e^- e^+ \rightarrow ZAA$ can be considered same. Therefore, the cross-section as a function of center of mass of energy for both, becomes equal. The plots for different polarizations of beam are shown in Figure ~\ref{fig:6_5.5}. The scattering amplitudes for Feynman diagrams shown in Figure ~\ref{fig:7_5.6} and ~\ref{fig:8_5.7} are same. We consider two different fixed values of extra Higgs states i.e., $m_H$ = 175 $GeV$ and $m_H$ = 150 $GeV$ and compare their cross-section. From Figure ~\ref{fig:6_5.5}, it is clear that for $m_H$ = 175 $GeV$ the maximum cross-section of unpolarized beam is 0.062 $fb$ at 1 $TeV$ which decreases slowly with increase in energy. The maximum cross-section of right handed polarized electron and left handed polarized positron beam is around 0.151 $fb$. When $m_H$ =150 $GeV$, then the unpolarized cross-section reaches about 0.075 $fb$  and then decreases slowly with higher energy. Additionally $\sigma_{RL}$ = 0.183 $fb$, which shows that $RL$ scenario enhances the cross-section. From comparison of two mass values, it is clear that a value of 150 $GeV$ shows maximum cross-section as compared to 175 $GeV$.\\

\subsubsection{ $e^- e^+ \rightarrow AHh$ }

\begin{figure}
\centering     
\subfigure[]{\label{fig:a}\includegraphics[width=60mm]{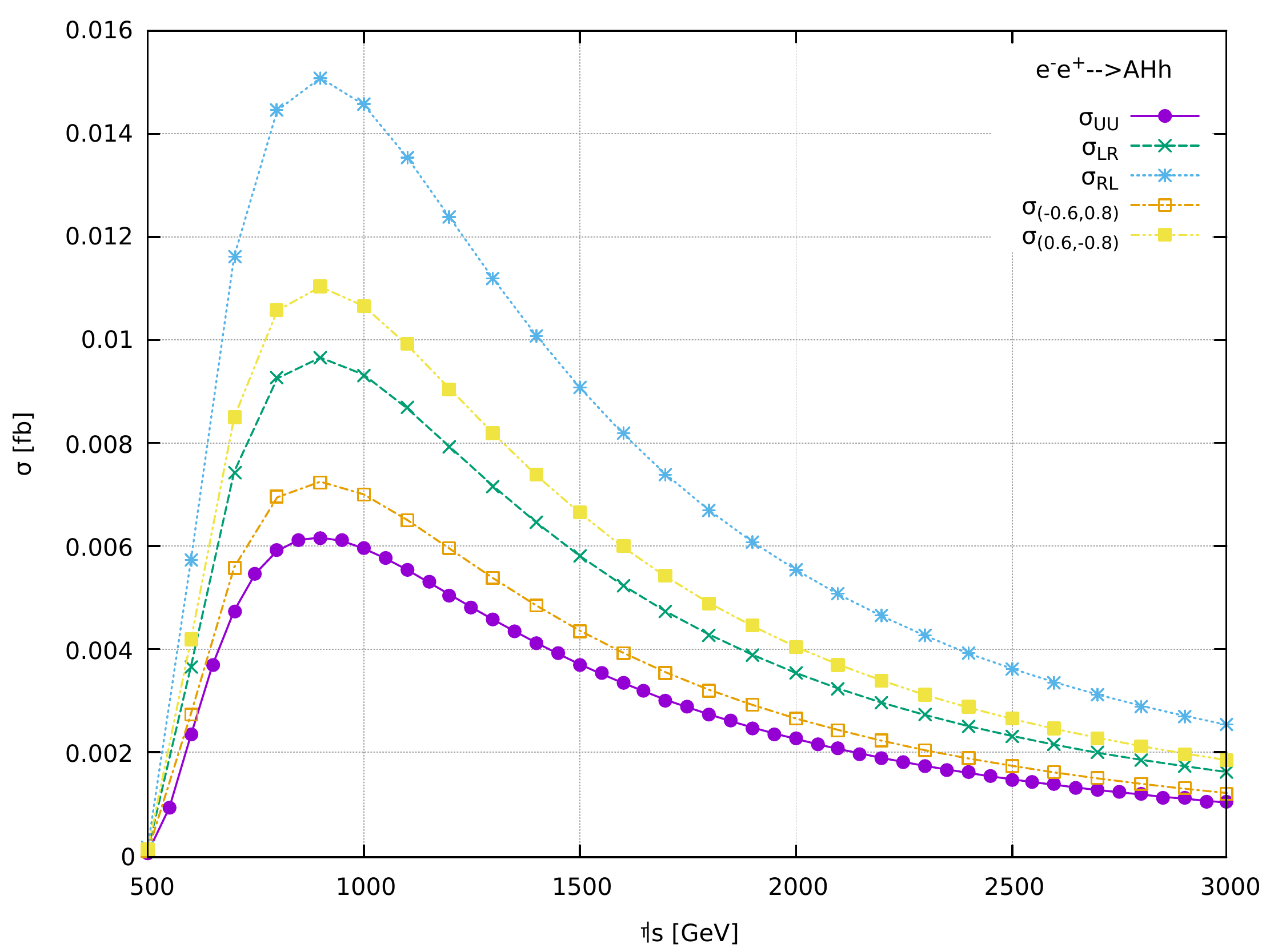}}
\subfigure[]{\label{fig:b}\includegraphics[width=60mm]{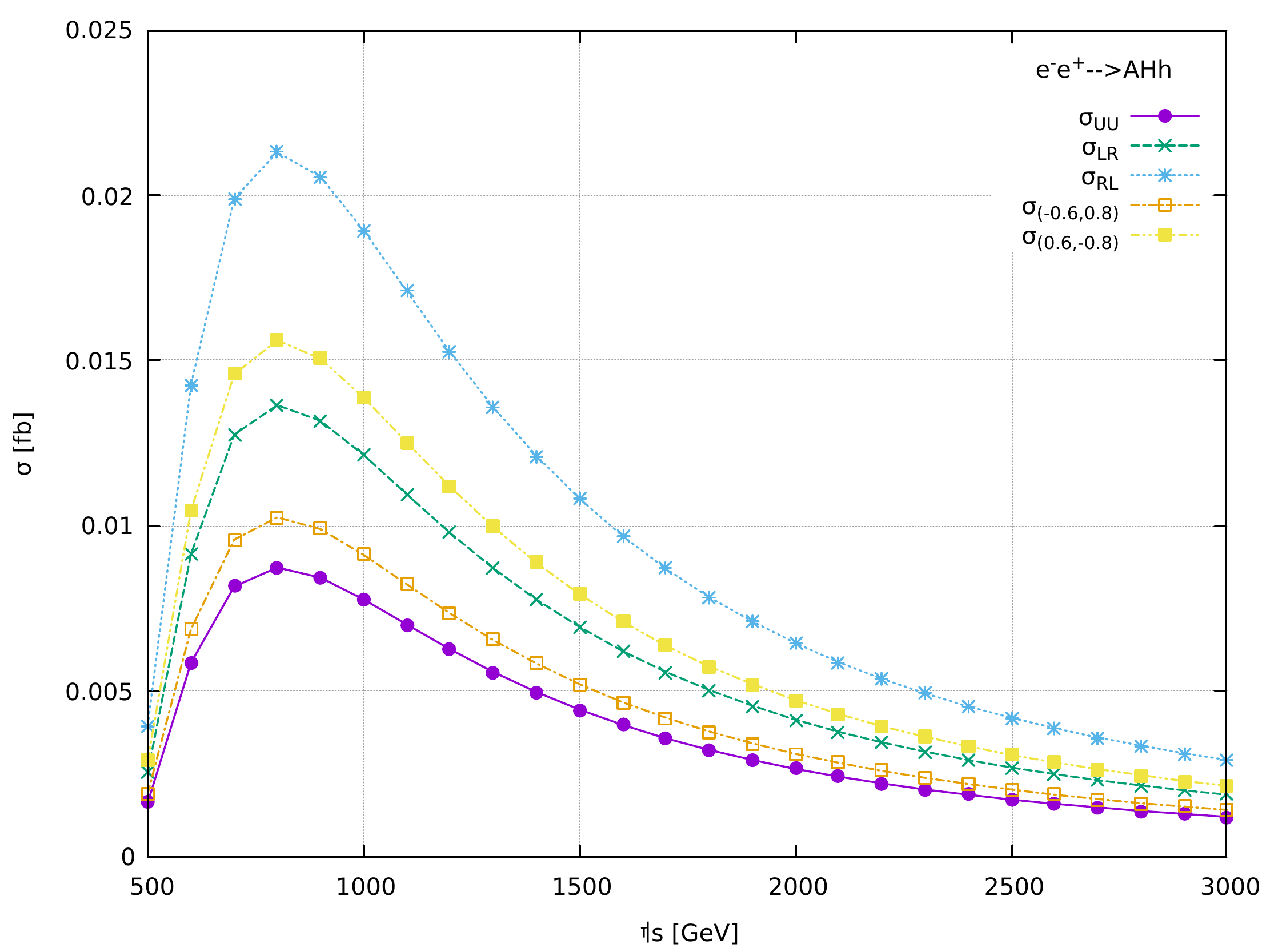}}
\subfigure[]{\label{fig:a}\includegraphics[width=60mm]{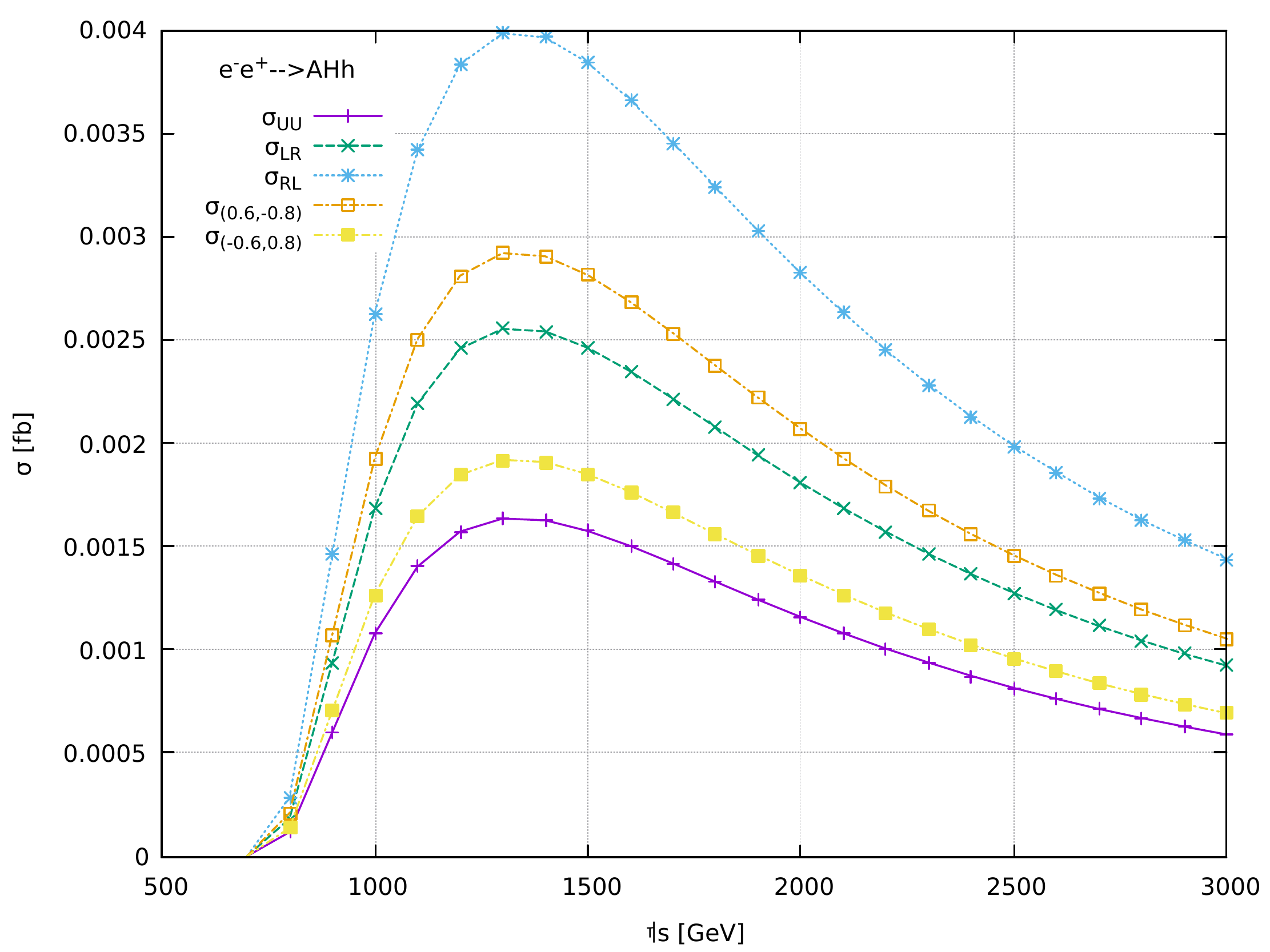}}
\subfigure[]{\label{fig:b}\includegraphics[width=60mm]{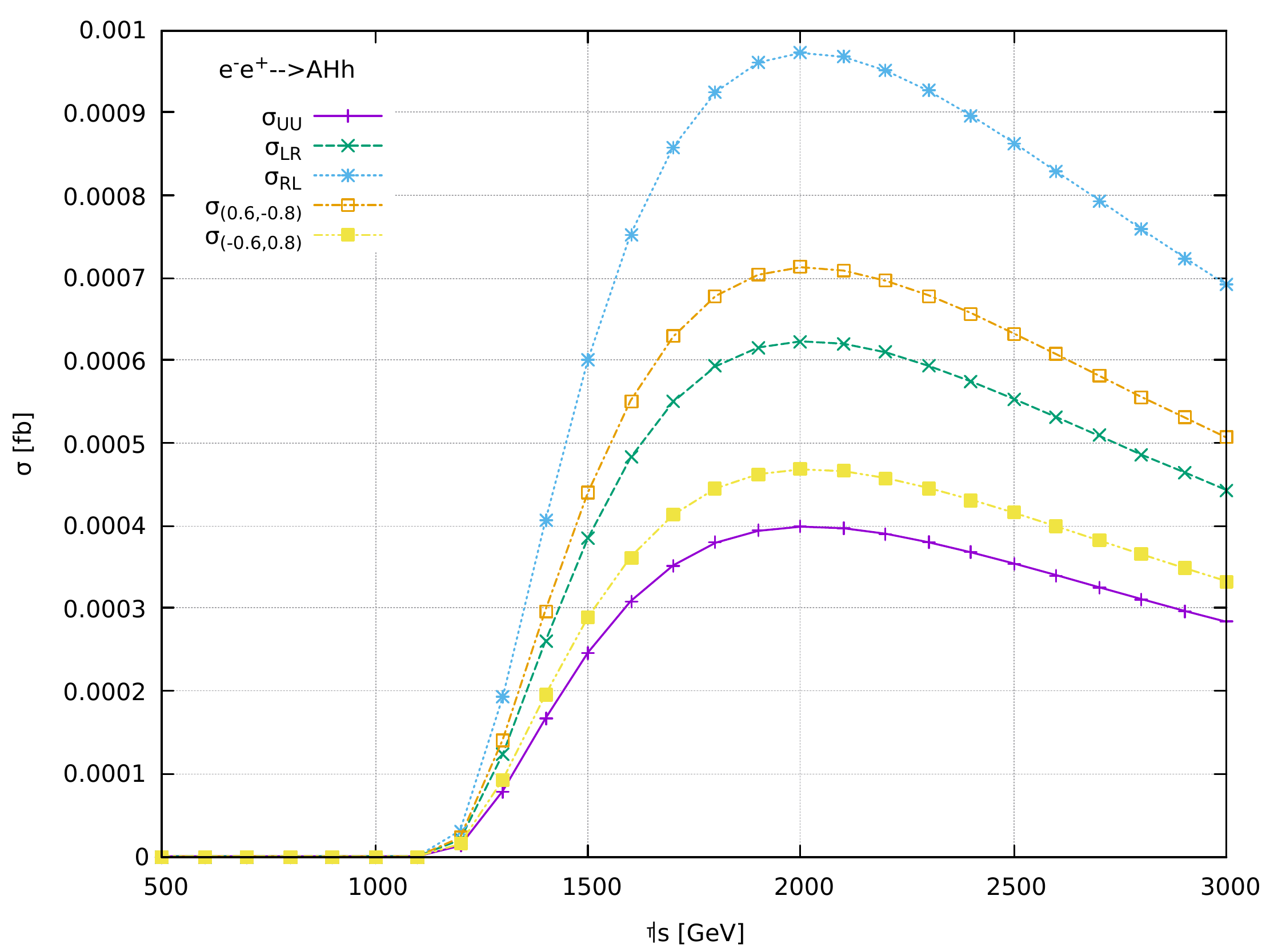}}
\caption{The distribution of cross-section $\sigma$ ($fb$) as a function of $\sqrt{s}$ ($GeV$) in 2HDM. The Higgs masses $m_H$ = $m_A$ = $m_{H^{\pm}}$ =175 $GeV$, $\tan \beta$=10 and $s_{\beta \alpha}$ = 1 are used (left). The Higgs masses are taken to be $m_H$ = $m_A$ = $m_{H^{\pm}}$ = 150 $GeV$,  $\tan \beta$=10 and $s_{\beta \alpha}$ = 1 (right).}
\label{fig:9_5.5}
\end{figure}

For the next scattering process $e^- e^+ \rightarrow AHh$, the distributions are shown in Figure ~\ref{fig:9_5.5}. When the mass of extra Higgs states are taken as 175 $GeV$ then the cross-section of $\sigma_{UU}$ (unpolarized) = 0.0059 $fb$ at $\sqrt{s}$=1 TeV. The maximum cross-section of polarized beam is $\sigma_{RL}$ = 0.0145 $fb$. Similarly when $m_H$ = $m_A$ = $m_{H^{\pm}}$ =150 $GeV$ are used then unpolarized cross-section is 0.0078 $fb$ and right-handed polarized electron beam and left-handed polarized positron beam ($\sigma_{RL}$) is 0.0189 $fb$. Coupling $g_{H^0 A^0 A^0}$ can be extracted from this scattering process but the cross-section is quite small. 

\begin{figure}[h]
    \centering
        \includegraphics[scale=0.50]{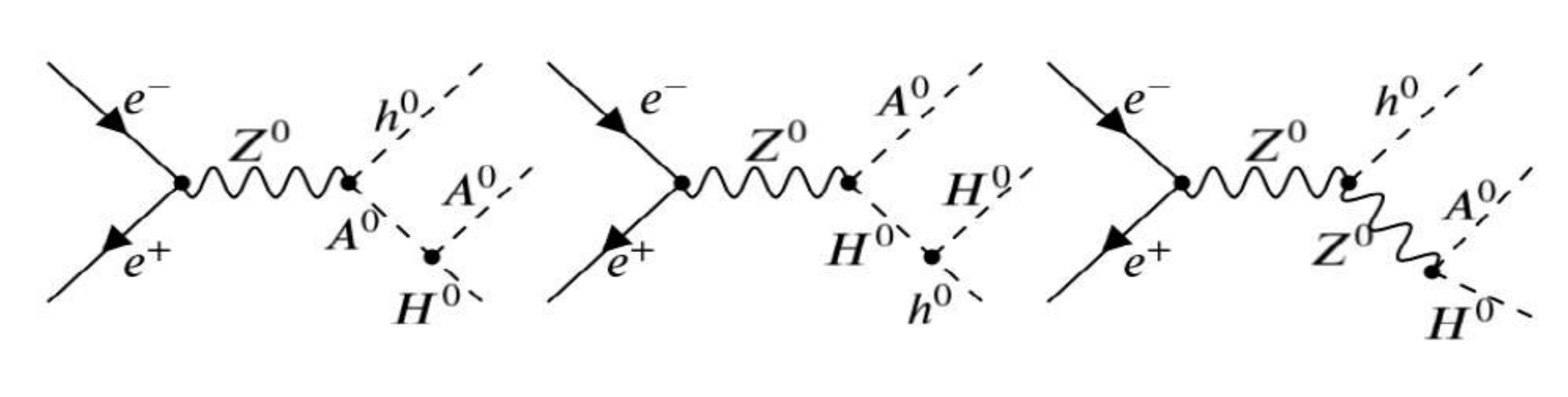}
    \caption{Feynman diagrams contributing to the scattering process   $e^- e^+ \rightarrow \text{ZHH}$ at ILC.}
    \label{fig:10_5.9}
\end{figure}

\subsubsection{$e^- e^+ \rightarrow AHH$}

\begin{figure}
\centering     
\subfigure[]{\label{fig:a}\includegraphics[width=60mm]{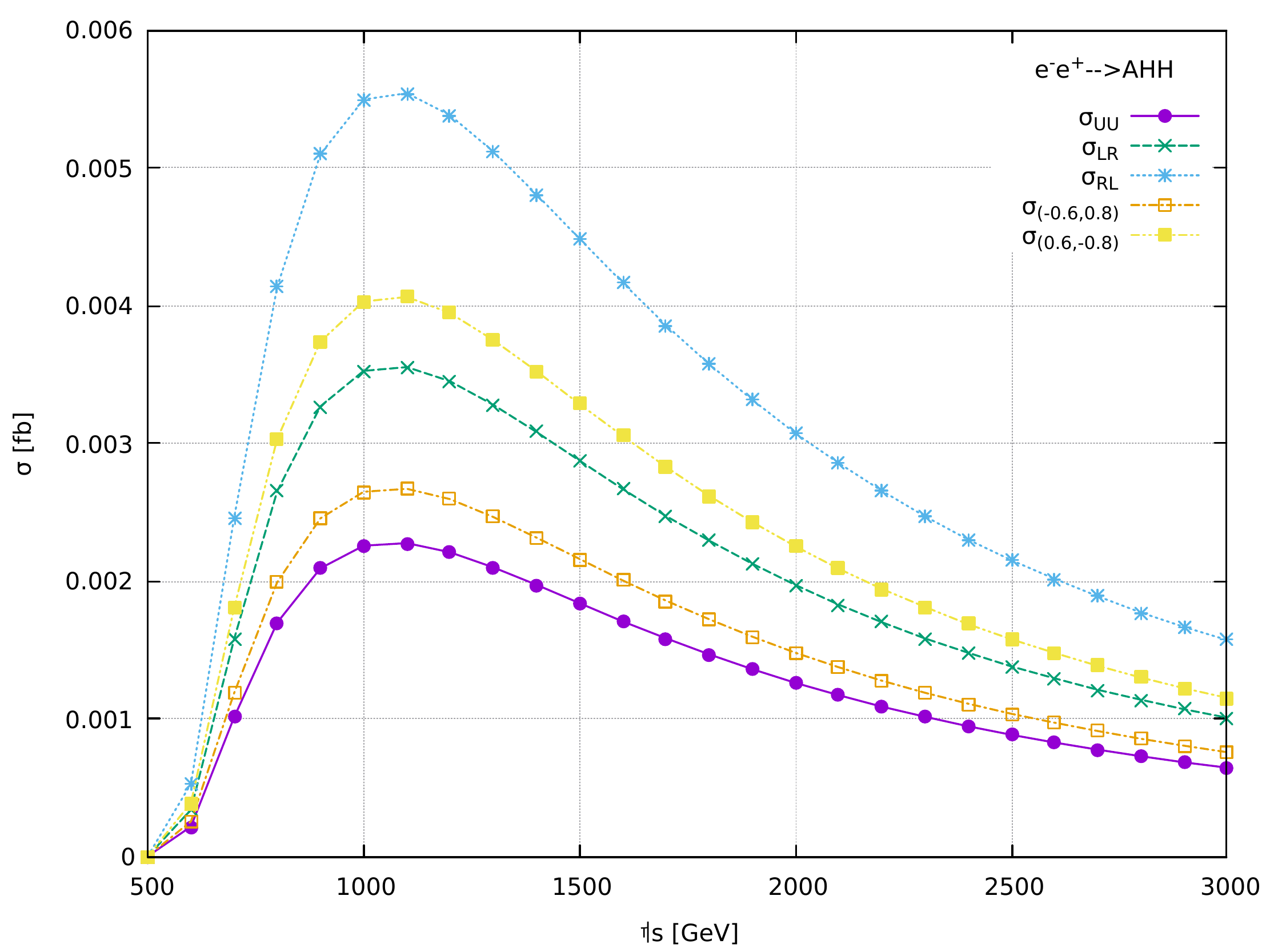}}
\subfigure[]{\label{fig:b}\includegraphics[width=60mm]{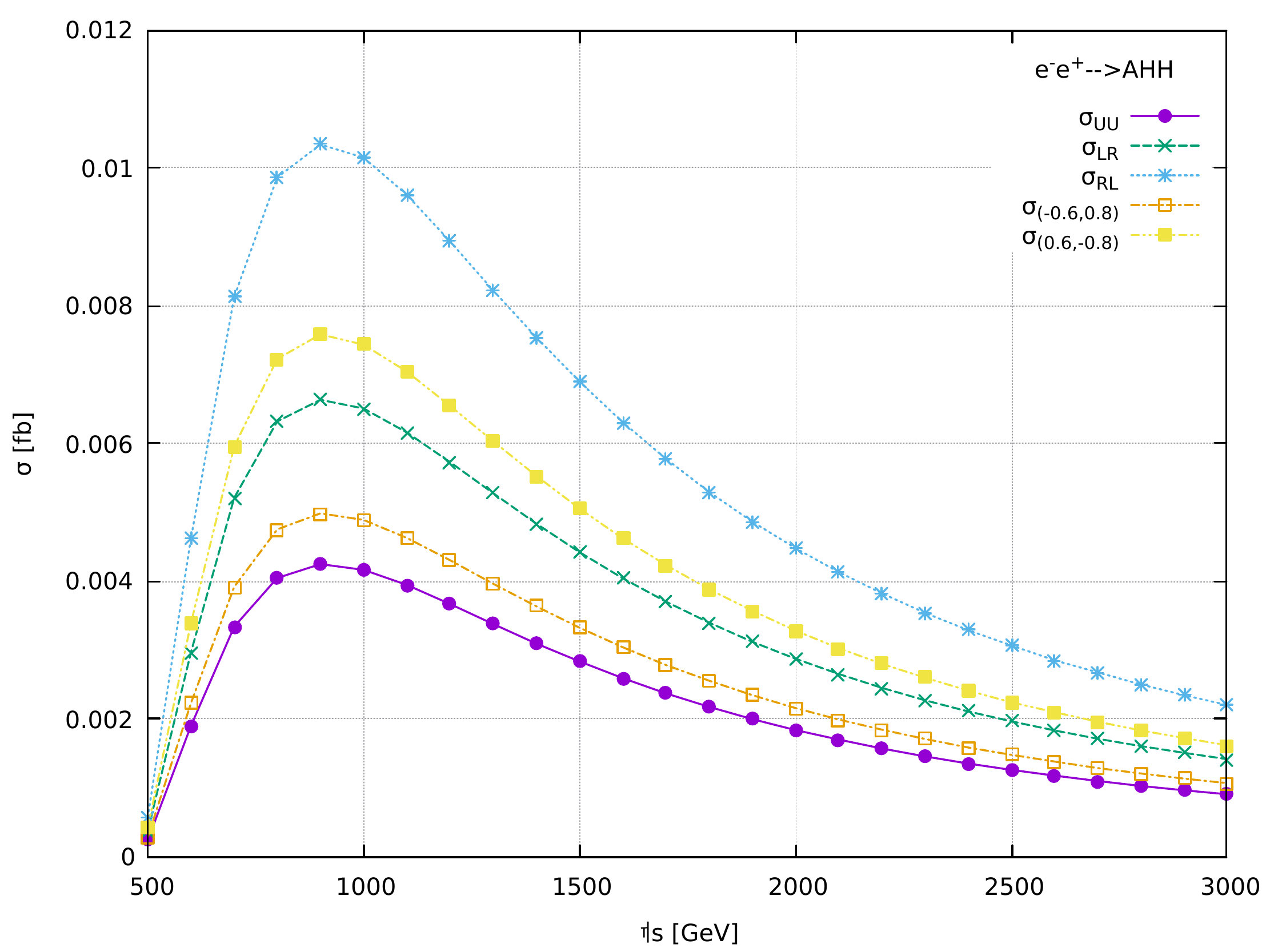}}
\subfigure[]{\label{fig:a}\includegraphics[width=60mm]{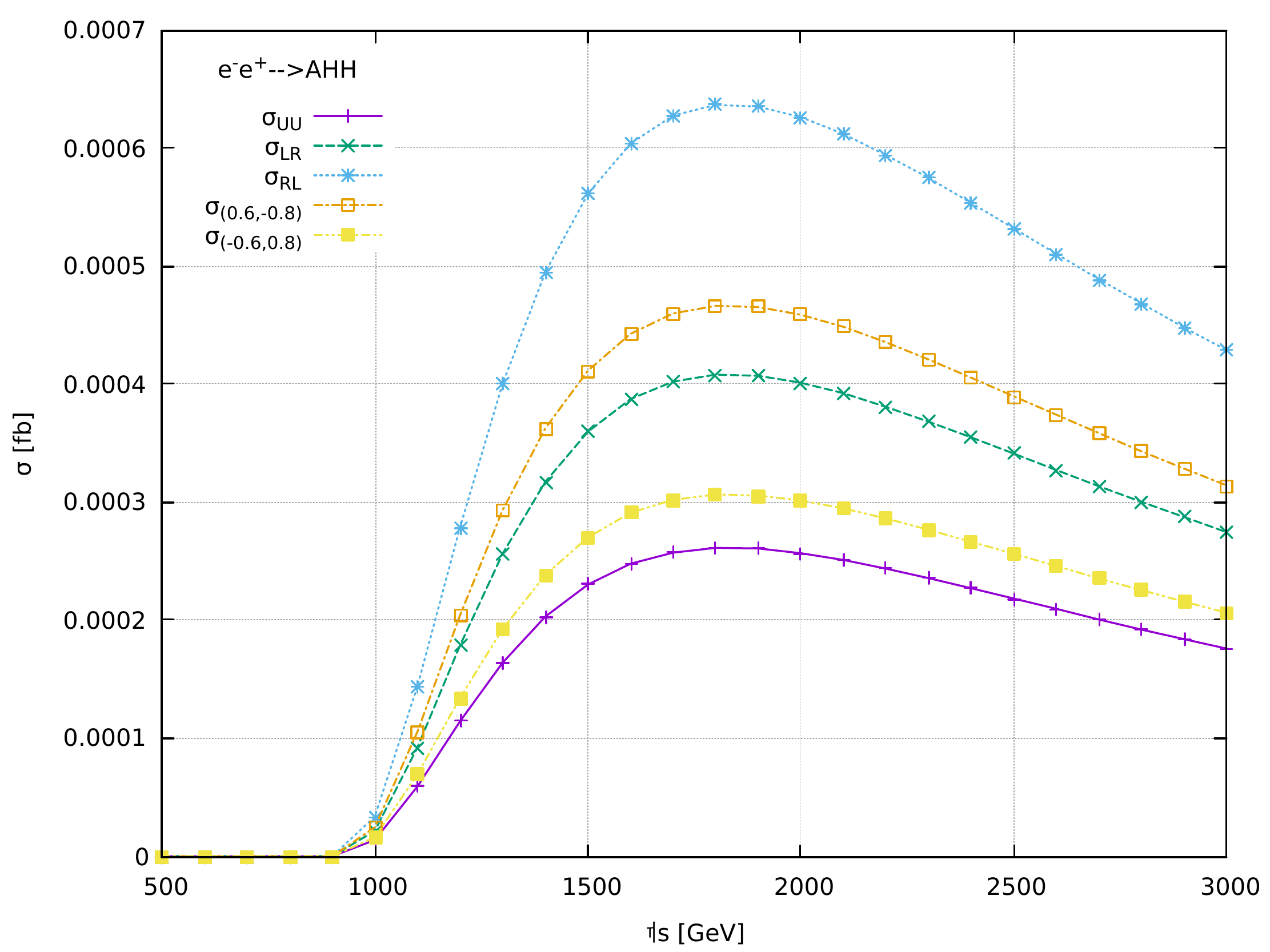}}
\subfigure[]{\label{fig:b}\includegraphics[width=60mm]{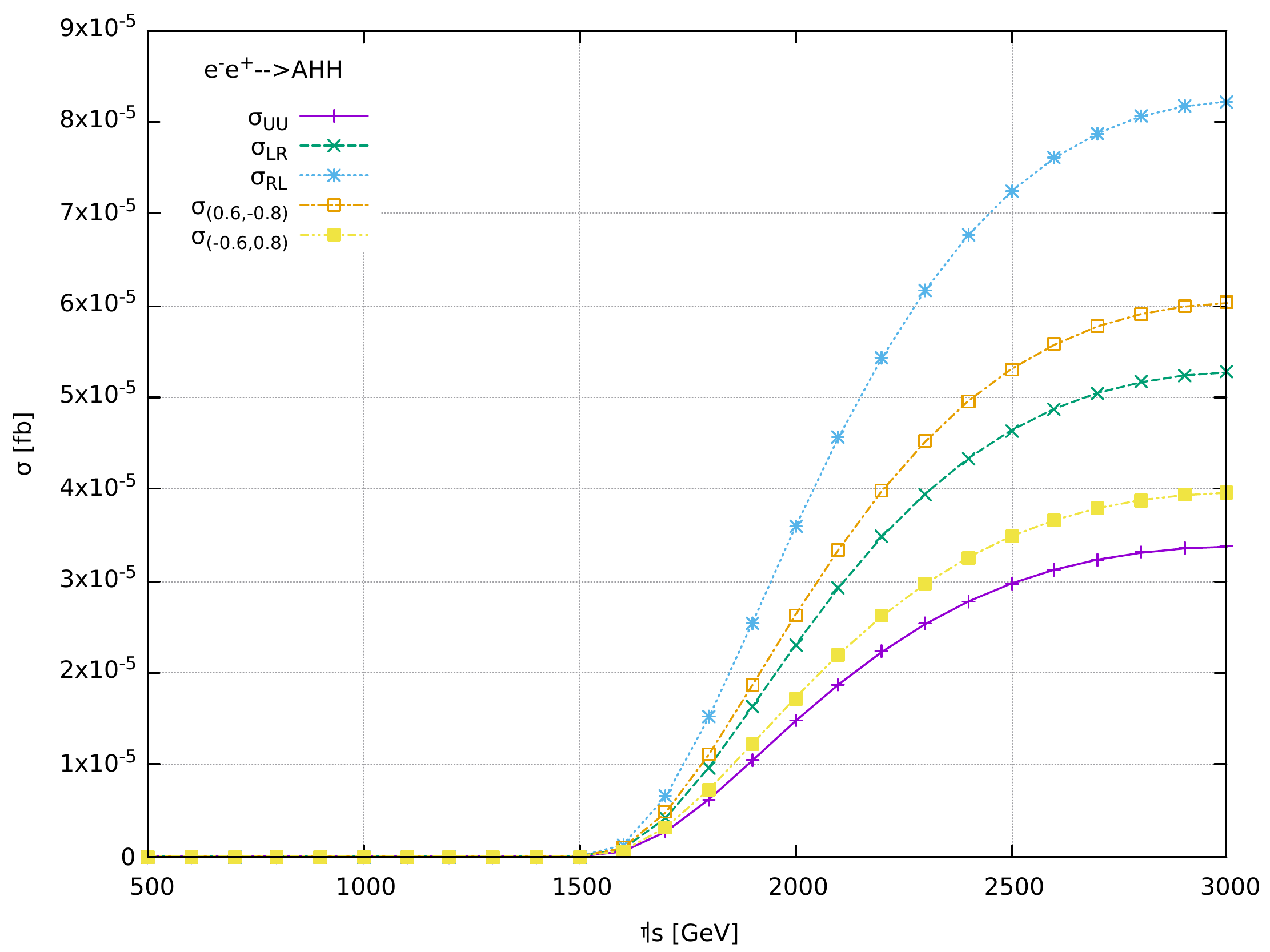}}
\caption{The distribution of cross-section $\sigma$ ($fb$) as a function of $\sqrt{s}$ ($GeV$) at 2HDM. The Higgs masses $m_H$ = $m_A$ = $m_{H^{\pm}}$ = 175 $GeV$, $\tan{\beta}$ = 10 and $s_{\beta \alpha}$=1 are used (left). The Higgs masses are taken as $m_H$ = $m_A$ = $m_{H^{\pm}}$ = 150 $GeV$, $\tan{\beta}$ = 10 and $s_{\beta \alpha}$=1 (right).}
 \label{fig:11_5.9}
\end{figure}

The last cross-section is calculated for the scattering process $e^- e^+ \rightarrow AHH$. The cross-section plot is shown in Figure ~\ref{fig:11_5.9}. This process has smallest cross-section as compared to others. The unpolarized beam has a cross-section of 0.0023 $fb$ at $\sqrt{s}$ = 1$TeV$, where $m_H$ = $m_A$ = $m_{H^{\pm}}$ = 175 $GeV$. In case of $m_H$ = $m_A$ = $m_{H^{\pm}}$  = 150 $GeV$ the cross-section of unpolarized beam is around 0.0042 $fb$ at 1$TeV$  which drops rapidly with increase in energy. 

\begin{figure}
\centering
\includegraphics[scale=0.50]{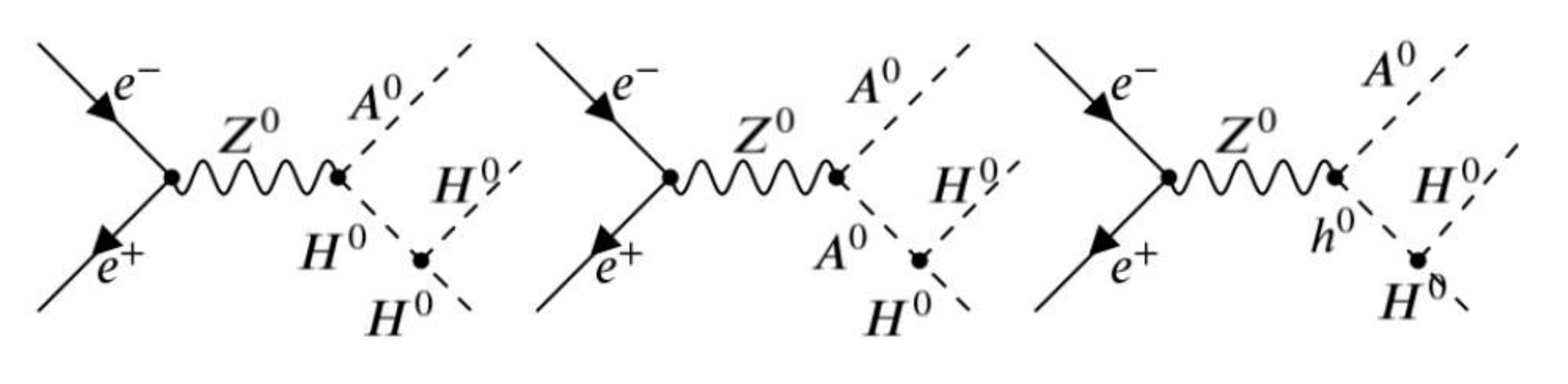}
\caption{The feynman diagram for scattering process $e^-e^+ \rightarrow$ AHH at ILC.}
\label{fig:12_5.10}
\end{figure}

\section{Dependance of $\tan{\beta}$ and $m_H$ On The Production Cross-section}

 \begin{figure}
\centering     
\subfigure[]{\label{fig:a}\includegraphics[width=60mm]{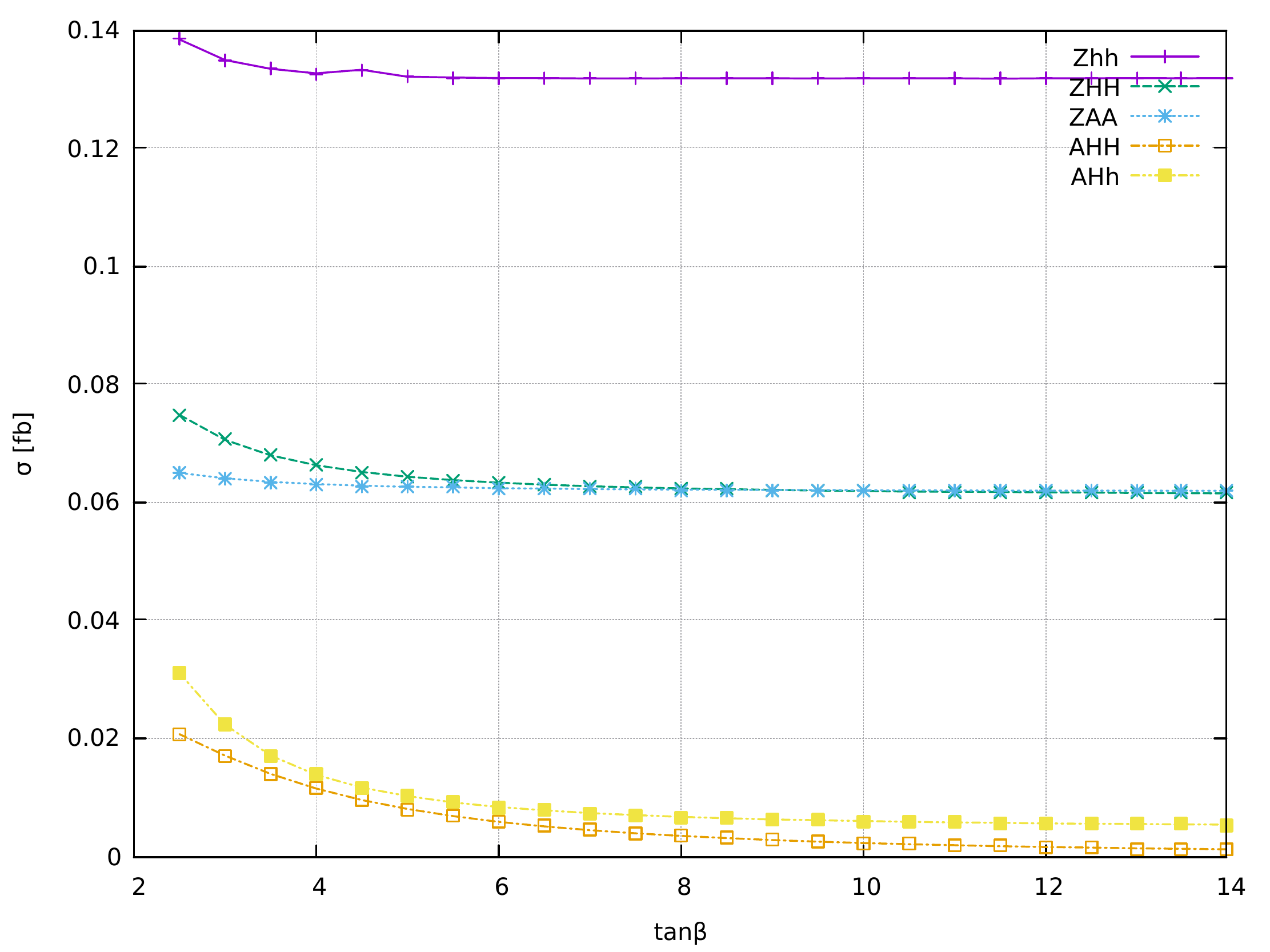}}
\subfigure[]{\label{fig:b}\includegraphics[width=60mm]{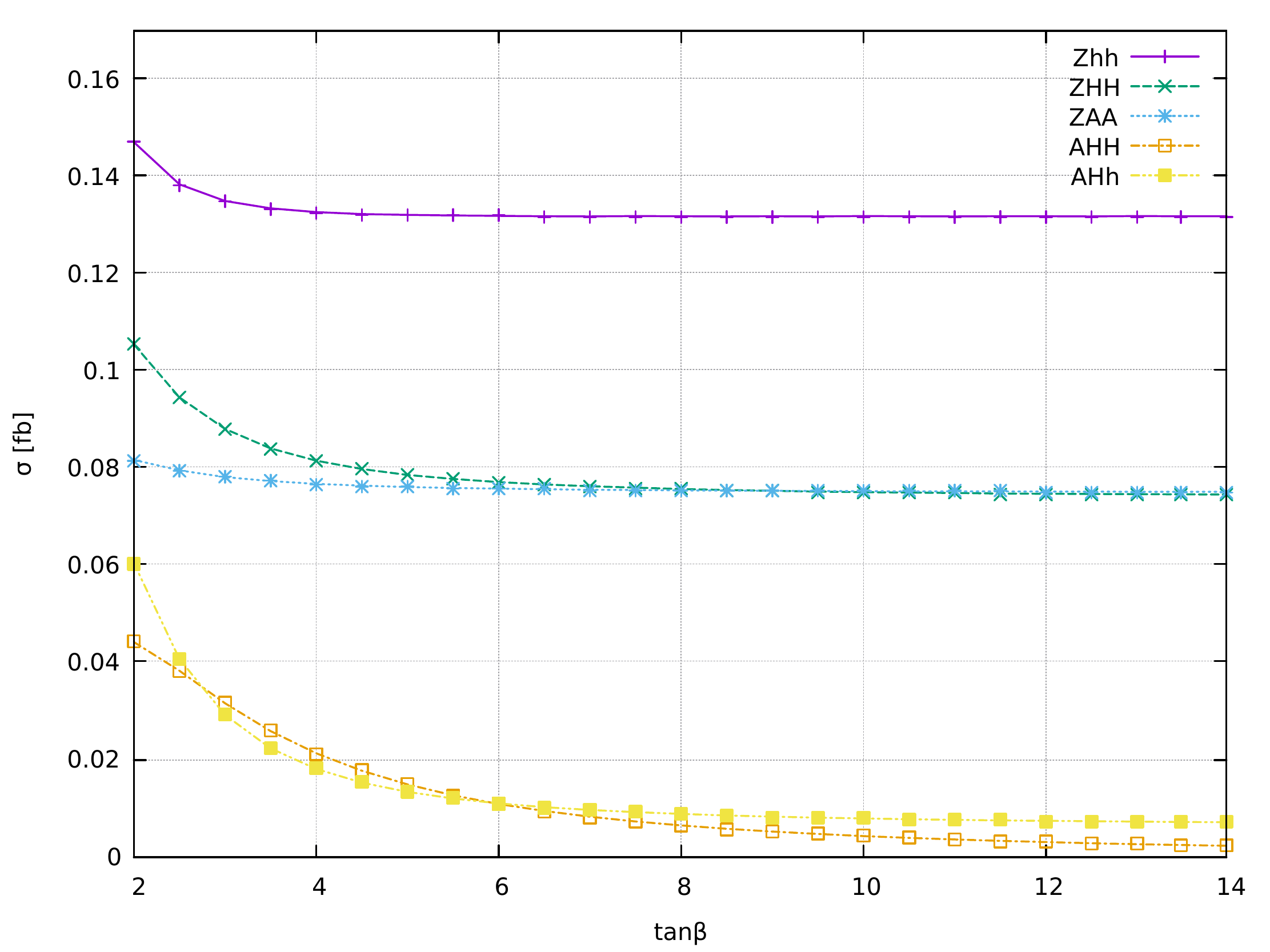}}
\caption{Cross-section distribution as a function of $\tan \beta$ for all the processes at $\sqrt{s}$ = 1 $TeV$. The values used are, $m_H$= 175 $GeV$ (left) and $m_H$ = 150 $GeV$ (right).}
 \label{fig:13_5.11}
\end{figure}

The variation of production cross-section for all processes, with change in $\tan{\beta}$ at $\sqrt{s}$ = 1$TeV$ is shown in Figure 5.11. It can be seen that for the process Zhh, production cross-section remains constant for the two mass values. The reason being the fact that, the coupling  $g_{h^0 h^0 h^0}$ for this process is the same as SM one. In the exact alignment limit, the production cross-section for processes ZHH and ZAA have the same distribution because both processes are a function of the same factors as given in Equations ~\ref{eq:4.5b} and ~\ref{eq:4.6b}. Next, the production cross-section of the processes AHH and AHh are maximum at the low value of $\tan{\beta}$ and decreases at higher values of $\tan{\beta}$. 
\begin{figure}[h]
    \centering
    \includegraphics[scale=0.40]{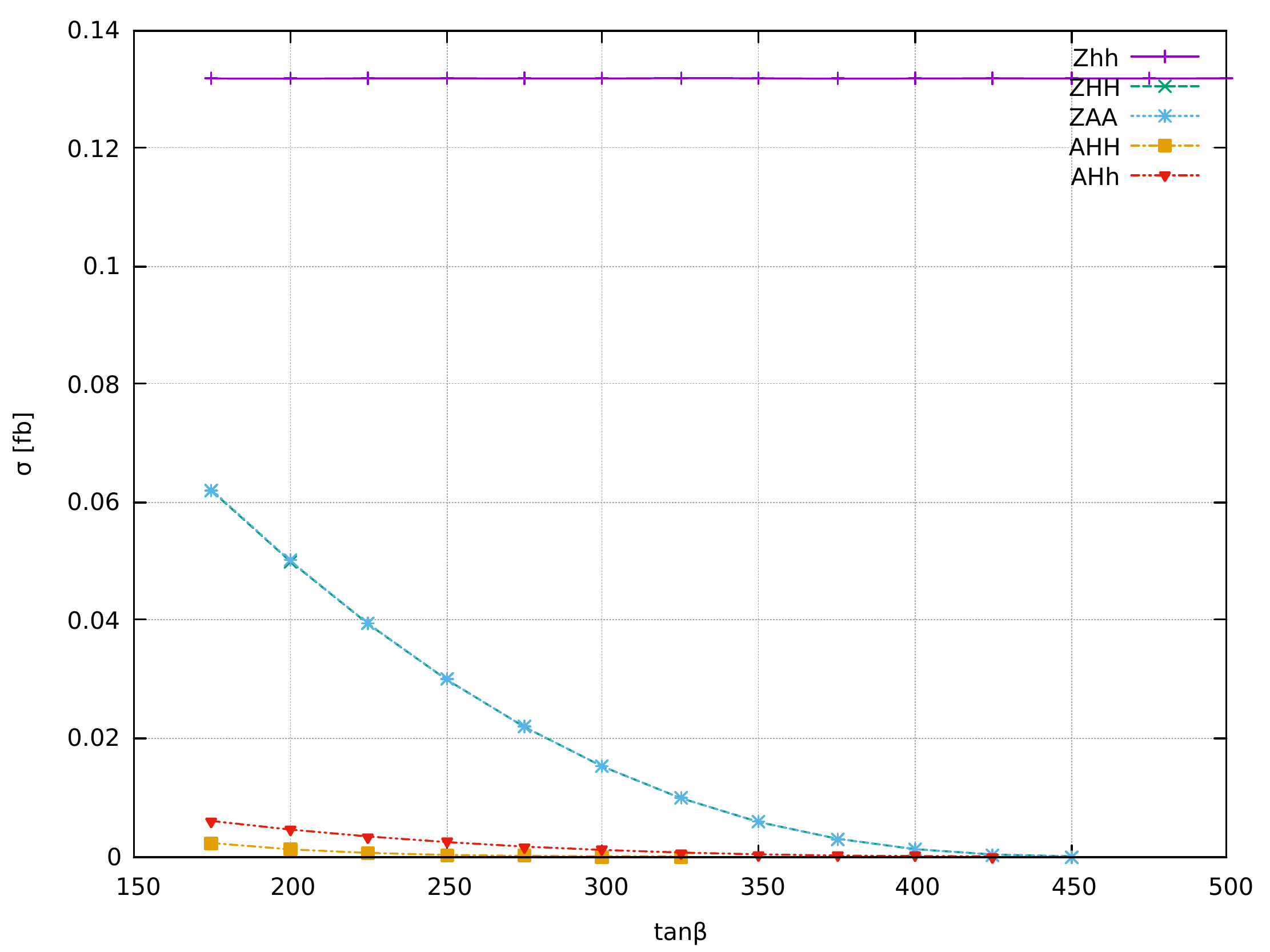}
    \caption{The distribution of cross-section as a function of tan$\beta$ at $\sqrt{s}$ = 1 $TeV$, where $s_{\beta \alpha}$ = 1 and $m_\phi$ = 150 GeV.}
    \label{fig:14_5.12}
\end{figure}

The distribution of cross-section as a function of $m_H$ is shown in Figure ~\ref{fig:14_5.12}. It can be seen that the cross-section for all processes, decreases with increasing values of $m_H$. This is as expected because, when the mass of all the extra Higgs states is increased the phase space becomes narrow for particles in the final state. The process $Zhh$ is an exception, for which the production cross-section remains constant and does not change with a change in the value of $m_H$. 

\section{Identifying the Processes}
In this section, possible decays of neutral Higgs boson are discussed, and possible colliders for analyzing each of the processes are perceived. The expected number of events and some possible background channels are also discussed.

\subsection{Decays Of Neutral Higgs Bosons}
The branching ratios and decay widths for neutral Higgs Bosons a are calculated using 2HDMC-1.7.0, whose parameters are explained in Table ~\ref{table:9a}. The branching ratios for Higgs particles $H^0$, $h^0$ and $A^0$ are given as a pie chart. Decay width is defined as the probability of occurance of a specific decay process within a given amount of time. The calculation involves determination of the decay width of neutral Higgs boson based on masses of particles involved and their relevant vertices. While the branching ratio of each Higgs boson remains stable. Further the decay channels for each of the Higgs bosons remain the same whereas the Higgs mass $m_H$ is changed.


\begin{figure}[h]
    \centering
    \includegraphics[scale=0.3]{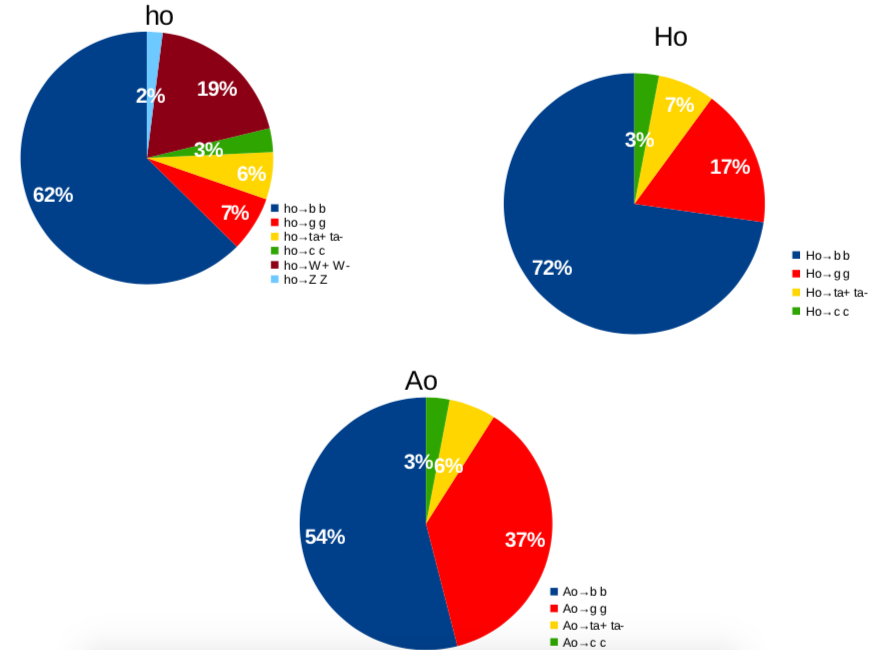}
    \caption{The branching ratios of all neutral Higgs boson decays for parameters defined in Table ~\ref{table:9a}.}
    \label{fig:15_5.14}
\end{figure}

From Figure ~\ref{fig:15_5.14}, it is clear that the prominent decay channel for all the neutral Higgs bosons is through b-quarks pair. The branching ratio for $h^0 \rightarrow b\bar{b}$ is $62 \%$. Similarly the branching ratios for $H^0 \rightarrow b\bar{b}$ and $A^0 \rightarrow b\bar{b}$ are $72 \%$ and $54 \%$ respectively. The other possible decay channels for $h^0$, $H^0$, and $A^0$ are g g and c c pairs. It is logical to say that, due to b-quark, gluon, and c-quark pairs the prominent pattern for each neutral Higgs boson is the di-jet final state in the detector. The next common decay channel is the $\tau \tau$. But the branching ratio of this process is small as compared to the di-jet signal. The $h^0$ also has other decay channels, which is through $W^+ W^-$ and $ZZ$ pair. Hence these decay channels can be considered as favorable but problem can arise due to leptonic decays of $W$-boson and extra $Z$-boson in the final state. Therefore the hadronic decay channels of $h^0$ boson promise more in the extraction of its pattern.

\subsection{Identification of processes and background channels}
In the previous section, decay channels of neutral Higgs boson are discussed and so acquiring the detector pattern of each channel is quite easy. The $Z$ boson decays through a hadronic, leptonic, and invisible channel. The branching ratios of $Z$ boson are $Z \rightarrow q\bar{q}$ ($\sim$0.70), $Z \rightarrow l \bar{l}$ ($\sim$0.10) and invisible($\sim$0.20), respectively. Hence there will be 4 $b$-quark jets +2 light jets (coming from $Z$ boson decay) in the final states. Unfortunately, the branching ratio of leptonic decay is small in comparison to the hadronic branching ratio, and there could be fewer events collected in the detector. Some possible patterns at the detectors, percentage of events and expected number of events are given in Tables ~\ref{table:10} and ~\ref{table:11} for an integrated luminosity values of $1ab^{-1}$ and $3ab^{-1}$.

\begin{table}[h!]
\begin{center}
 \begin{tabular}{|c|c|c|c|c|c|}
\hline
Detector patern & $Z^0 A^0 A^0$ & $Z^0 H^0 H^0$ & $Z^0 h^0 h^0$ &  $Z^0 H^0 h^0$ &  $Z^0 H^0 H^0$ \\
\hline
4b-quarks jets+2jets & 20.41 & 36.28 & 26.91 & 28.31 & 35.76 \\
6 jets & 60 & 58 & 37.30 & 62 & 78.71 \\
\hline
\end{tabular}%
\caption{The percentage of events for different decay channels.}
\label{table:10}
\end{center}
\end{table}

\begin{table}[h!]
\begin{center}
 \begin{tabular}{|c|c|c|c|c|c|c|c|c|c|c|}
\hline
Detector patern & \multicolumn{2}{c|}{ $Z^0 A^0 A^0$ } & \multicolumn{2}{c|}{ $Z^0 H^0 H^0$ } & \multicolumn{2}{c|}{ $Z^0 h^0 h^0$ } & \multicolumn{2}{c|}{ $Z^0 H^0 h^0$ } & \multicolumn{2}{c|}{ $Z^0 H^0 H^0$ } \\
\cline{2-11}
    & $1ab^{-1}$ & $3ab^{-1}$ & $1ab^{-1}$ & $3ab^{-1}$ & $1ab^{-1}$ & $3ab^{-1}$ & $1ab^{-1}$ & $3ab^{-1}$ & $1ab^{-1}$ & $3ab^{-1}$ \\
\hline
4b-quarks jets+2jets & 12 & 39 & 22 & 66 & 35 & 105 & 2 & 4 & 0.8 & 2.4 \\
6 jets & 38 & 114 & 37 & 111 & 48 & 144 & 3 & 9 & 1.7 & 5.1 \\
\hline
\end{tabular}%
\caption{Expected number of events for integrated luminosity of  $1ab^{-1}$ and $3ab^{-1}$.}
\label{table:11}
\end{center}
\end{table}

It is assumed that International Linear Collider can achieve a total integrated luminosity of $1ab^{-1}$ and $3ab^{-1}$ in its lifetime. The number of events that are expected for various processes in that case are given in Table ~\ref{table:11}at $\sqrt{s}$ = 1 $TeV$. Where the $m_H$ =175 $GeV$ and $\tan{\beta}$ = 10 are used. Unfortunately, some events for scattering process $A^0H^0h^0$ and $A^0H^0H^0$ are $\leq$10 and $\leq$ 6 with $3ab^{-1}$. The number of events could be increased by using polarized incoming beams. However, only this may not give enough cross-section to be useful for the measurement of these two processes.
There is some background signal which can hide the actual processes. In SM the most relevant and expected background channels are $e^+ e^- \rightarrow Zb\bar{b}b\bar{b}$, $e^+ e^- \rightarrow Zcccc$ and $e^+ e^- \rightarrow ZZ \rightarrow b\bar{b}b\bar{b}$. Hence reconstruction of the Higgs mass in each event could be beneficial. If neutral Higgses do not decay through bb pairs, then such b-quarks pair will give an invariant mass value that lies outside of the Higgs mass range and such events can be excluded. If the efficiency of b-tagging is measured around $80 \%$ or higher, and we require $H_i H_j \rightarrow b\bar{b}b\bar{b}$ + hadronic decay, this offers a definite pattern at the detector which is 4 b-tagged jets +2 light jets. A notable amount of events can be obtained from this pattern, whereas a big fraction of the background signal could be eliminated, because of b-tagged jet requirements. A Monte Carlo simulation of each of the processes is required to evaluate trigger efficiency and acceptance of the detector for different decay channels.

\section{Conclusion}
In this study, the production cross-section of various scattering processes is calculated in $e^+e^-$ collider. The scattering process selected to determine the triple Higgs-self coupling is governed by 2HDM. The 2HDM is examined by considering experimental constraints on charged Higgs boson. These constraints probe the exact alignment limit $s_{\beta \alpha}$ = 1. There are eight possible Higgs self-couplings, two out of which, the ones for charged Higgs, are not discussed in this study. Due to exact alignment limit $s_{\beta \alpha}$ = 1, one of the Higgs self-coupling $g_{h^0 h^0 H^0}$ vanishes, therefore only five Higgs self-coupling survive out of eight. The contribution of Higgs self-coupling for each scattering process is described in Table 5.1. The coupling could be determined using the process $Z^0 h^0 h^0$ which is the same as in Standard Model. The final state of the process $Z^0 A^0 A^0$ and $Z^0 H^0 H^0$ helps in extraction of coupling $g_{h^0 A^0 A^0}$ and $g_{h^0 H^0 H^0}$. These processes have an acceptable cross-section. Similarly the coupling $g_{H^0 A^0 A^0}$ is obtained by the process $e^+e^- \rightarrow A^0 h^0 h^0$. But the cross-section is in atto-barn, and might not be enough for accumulation of sufficient number of events. At last, the process $e^+e^- \rightarrow A^0 H^0 H^0$ extracts the Higgs self-coupling $g_{H^0 H^0 H^0}$ with the help of $g_{H^0 A^0 A^0}$, but only if it could ever possibly be determined through the process $A^0 H^0 H^0$. Although, examining the previous process could be difficult, calculating the coupling $g_{H^0 H^0 H^0}$ is also a challenge using $A^0 H^0 H^0$ process. The calculation of production cross-section for all the above scattering processes, both with and without polarized beams shows that the incoming polarized beam enhance the cross-section. The right-handed electron beam and left-handed positron beam enhanced the cross-section up to a factor of 2.4 for all processes. The calculation is carried out in exact alignment limit $s_{\beta \alpha}$ = 1 and $m_H$ = 175 $GeV$ and $m_H$ = 150 $GeV$. The calculation shows that the cross-section increases when $m_H$ =150 $GeV$ is used in comparison to 175 $GeV$, for all processes, except $Z^0 h^0 h^0$.
The decay widths (not given) and branching ratios of neutral Higgs bosons are also calculated. The study shows that all neutral Higgs bosons have identical decay channels for the specific choice of parameters. The dominant decay channel of all neutral Higgs bosons is through $b\bar{b}$ pair, gluon pair, $c\bar{c}$ pair and with a small fraction of $\tau\bar{\tau}$.
For the measurement of cross-section distribution, ILC has the biggest potential for contribution as compared to the proposed all other future lepton colliders. ILC will operate at a centre of mass energy ranging up to 1TeV. The Future Circular Collider can also make measurements and compete for the couplings $g_{h^0 h^0 h^0}$, $g_{h^0 A^0 A^0}$ and  $g_{h^0 H^0 H^0}$ by operating at a centre of mass energy range up to 0.5 $TeV$. However, the Circular Electron-Positron Collider does have sufficient centre of mass energy to investigate the Higgs self-coupling even for a process like ZHH  in SM.
The invention of the Higgs boson at LHC has established the Higgs mechanism, generating mass to all particles. This was the last piece of SM and further no clue to new physics has been observed. The simplest extension of SM is 2HDM. This study shows the possible measurement of trilinear Higgs-coupling in the future lepton colliders, which plays a vital role in reconstructing Higgs potential.\\

\end{document}